\documentclass[useAMS, usenatbib]{mn2e} 
\usepackage{times}
\usepackage{aas_macros}
\usepackage{graphicx}
\title[{\small CODEX} II: Properties of 4 M-Type Model series]{Dynamical Opacity-Sampling Models of Mira
  Variables. II: Time-Dependent Atmospheric Structure and Observable Properties of 4 M-Type 
Model Series.} 

\author[M.J. Ireland et al.]{M.J. Ireland\thanks{michael.ireland@mq.edu.au}$^{1,4,5}$, M. Scholz$^{1,2}$, P.R. Wood$
^3$ \\
$^1$Sydney Institute for Astronomy (SIfA), School of Physics, University of Sydney NSW
 2006, Australia\\
$^2$Zentrum f\"ur Astronomie der Universit\"at Heidelberg (ZAH),
  Institut f\"ur Theoretische Astrophysik, Albert-Ueberle-Str.2, \\69120 Heidelberg, Germany\\
$^3$Research School for Astronomy and Astrophysics, Australian National 
University, Canberra ACT 2600, Australia\\
$^4$Department of Physics and Astronomy, Macquarie University, NSW 2109, Australia\\
$^5$Australian Astronomical Observatory,  PO Box 296, Epping, NSW 1710, Australia}
\begin{document}

\pagerange{\pageref{firstpage}--\pageref{lastpage}} \pubyear{2008}

\maketitle

\label{firstpage}

\begin{abstract}
 We present 4 model series of the {\small CODEX} dynamical
 opacity-sampling models of Mira variables with solar abundances, designed to have
 parameters similar to $o$~Cet, R~Leo and R~Cas. We demonstrate that
 the {\small CODEX} models provide a clear physical basis for the molecular shell
 scenario used to explain interferometric observations of Mira
 variables. We show that these models generally provide a good match
 to photometry and interferometry at 
 wavelengths between the near-infrared and the radio, and make the model
 outputs publicly available. These model also demonstrate
 that, in order to match visible and infrared observations, the
 Fe-poor silicate grains that form within 3 continuum radii must have small
 grain radii and therefore can not drive the winds from O-rich Mira variables.
\end{abstract}

\section{Introduction} 

Asymptotic Giant Branch (AGB) stars represent the final fusion-powered
stage in the evolution of solar-type stars, and the engine by which the vast
majority of the material in our Galaxy is recycled from stars back to
the interstellar medium \citep{Gail03}. Mira variables represent the
final stage in AGB 
evolution before they become dust-enshrouded and difficult to observe.
They are so bright in the infrared that they can be used as
competitive extragalactic distance indicators and probes of star
formation history \citep{Rejkuba04,Menzies10}.
They are also unique amongst stellar classes in their opportunity for detailed
observations: light curves that differ in shape and amplitude in
different bandpasses, photospheres that show different sizes and
structure as a function of wavelength, velocity-resolved motions and
complex spectra.

The observational literature on Mira variables is very
extensive, as partly detailed below in
Section~\ref{sectInterferometry}. In order to make sense of these
observations, comprehensive models are required that link physical
parameters to pulsation, and pulsation to observed properties and
mass-loss. The previous generation of models
\citep[e.g.][]{Hofner98,Hofmann98} suffered from grey or mean-opacity 
like approximations in their radiative transfer codes, so were not
ideally suite to interpreting many observed properties, for example
visible brightness or high-resolution spectroscopy. The next
generation of models (Upsalla:\citealt{Hofner03},{\small
  CODEX}:\citealt{Ireland08}) are better suited for modern
observational comparisons, but extensive grids
have not yet been produced, both because of solvable but difficult
computational issues and the lack of a clear calibration for model
parameters. 

Here we present four physical model series for M-type Mira
variables, as a first step in tuning and testing the {\small CODEX}
models to derive physical parameters of Mira variables from
observations, and to gain physical insight into the dominant physical
processes in Mira-variable atmospheres. Observational predictions
including predictions for infrared interferometry of the model series
are made available online so that new observations can easily be
compared to these models.

One of these model series has parameters based on $o$~Cet, 
one based on $R$~Leo and two are based on R~Cas with different
assumptions. For all models, only the pulsation period is guaranteed
to match observations, and in this paper we aim to examine the other
model outputs in order to more closely target model series to real
stars in future papers. 
The model construction was 
described in Paper~I \citep{Ireland08}: they begin with input
parameters of mass, 
luminosity and composition, with three other free parameters being
microturbulent velocity, mixing length ($\alpha_m$) and turbulent
viscosity ($\alpha_\nu$). Pulsation is self-excited (i.e. it occurs
spontaneously),  
and the temperature of all layers is calculated by solving the
conservation of energy equation via a detailed opacity-sampling
method. 

Details of the parameter choices for the model series are
given in Section~\ref{sectMethods}, as well as basic comparison of the
model light curves to observations.
Model predictions for spectra, in particular the effect of extension
on spectra, are tested in Section~\ref{sectSpectra}, and the
models are compared to observations of $o$~Cet including infrared and
radio interferometry in Section~\ref{sectInterferometry}.  In
Section~\ref{sectShell}, we compare the model 
structures to previously published ad-hoc molecular shell papers, and
in Section~\ref{sectRadAccel} we discuss the mass loss rates of the
models and the driving mechanisms. In Section~\ref{sectParams} we
discuss the effect of input parameters on the models, and the
possibility for better calibrating the input parameters so that,
e.g. the mass of individual Miras could be inferred from
models. Finally, in Section~\ref{sectConclusions} we conclude and
discuss plans for future work. 

\section{Model Parameters and Description}
\label{sectMethods}

A detailed description of the model construction was given in
Paper~I. Briefly, the models consist of self-excited grey models
that determine the atmospheric pressure stratification
and luminosity. The temperature profile is then re-iterated using an
opacity sampling code with 4300 wavelength points, assuming radiative
and local thermodynamic equilibrium. Dust formation follows
\citet{Ireland06}, except that we drop the Rayleigh approximation,
instead replacing it by a smooth fit to the Mie approximation of
spherical grains,
weighting the Rayleigh scattering by:

\begin{equation}
 \sigma_M(a,\lambda) = \sigma_R(a,\lambda) (1 + 4.5(\frac{a}{\lambda})^4)^{-1},
\end{equation} 

where $a$ is the grain radius, $\lambda$ the wavelength of radiation,
$\sigma$ the scattering cross-section and $\sigma_R$ the scattering
cross-section in the Rayleigh approximation. This cross-section
$\sigma_M$ is the total cross-section weighted by (1-$\mu$), where
$\mu$ is the impact parameter. This weighting ensures that the
radiative acceleration on dust is correct in the presence of
forward-scattering.

The choice of free parameters was only briefly 
discussed in Paper~I in the context of the {\tt o54} model series, 
based on the parameters of the prototype Mira variable $o$~Cet. The input parameters
for all 4 model series presented here are given in
Table~\ref{tableInputParams}, and the reasons for their choice give in
sub-sections below.

The
behavior over $10^4$ days of each nonlinear pulsation model series is shown in
Figures~1--4.     
As each model series runs for many cycles, we chose only a
few typical cycles for detailed examination.  In each of these cycles, 
$\sim$10 representative models were extracted and their velocity 
and pressure structures used as input to the model atmosphere code which,
after temperature iteration, give spectra and
centre-to-limb variation (CLV) for the models.  The cycles during which
models were extracted are shown as shaded regions in Figures~1--4
and the actual models extracted are shown as circles in the top panels of
these figures.
The instantaneous physical parameters and shock front locations for the
chosen models are given in Tables~\ref{tableo54_hx}--\ref{tabler81}
(available in their entirety in the on-line 
version of the journal). We note that the model radius is defined in these
tables as the radius where the Rosseland optical depth is unity, and
the effective temperature is defined by this radius. 

\begin{table}
\caption{Parameters of 4 model series. The mass $M$, luminosity $L$,
  metallicity $Z$, mixing-length parameter $\alpha_m$ and turbulent
  viscosity parameter $\alpha_\mu$
  are input parameters, and the parent-star radius $R_p$
  and period $P$ are derived parameters.}
\begin{tabular}{llrrrrrrr}
Name & M & L & Z & $\alpha_m$ &$\alpha_\nu$& $R_p$ & $P$ & $P_{\rm linear}$\\ 
\hline
{\tt o54} & 1.1  & 5400 & 0.02 & 3.5 & 0.25 & 216 & 330 & 261 \\
{\tt R52} & 1.1  & 5200 & 0.02 & 3.5 & 0.25 & 209 & 307 & 243 \\
{\tt C50} & 1.35 & 5050 & 0.02 & 2.0 & 0.24 & 291 & 427 & 408 \\ 
{\tt C81} & 1.35 & 8160 & 0.02 & 3.5 & 0.32 & 278 & 430 & 374 \\
\end{tabular}
\label{tableInputParams}
\end{table}

\begin{figure}
\vspace{-3mm}
\includegraphics[scale=0.47]{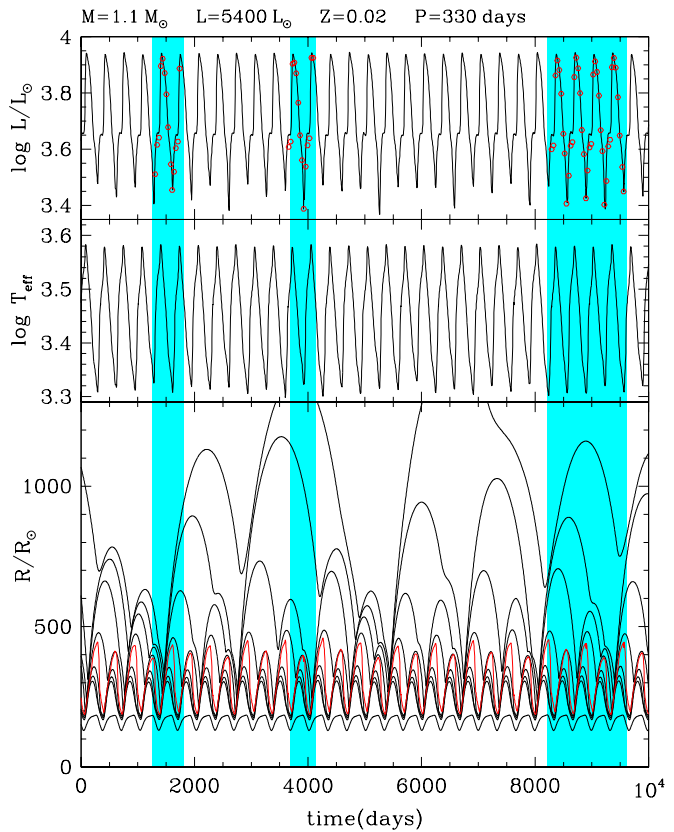}
\caption{The luminosity (top panel), effective temperature (middle panel), and
the radii of a representative selection of mass zones (bottom panel)
plotted against time for model {\tt o54}. The red line in the
bottom panel shows the position of the point where the grey
approximation optical depth
$\tau_g = \frac{2}{3}$.  $T_{\rm eff}$ in the middle panel is here
defined as the temperature where $\tau_g=2/3$,
which is close to the effective temperature $\propto (L/R^2)^{1/4}$ of the
non-grey atmospheric stratification. The shaded regions show
the time intervals in which models were selected for detailed
atmospheric model computation.  The selected models are circled in the top
panel.  The mass, luminosity, metallicity and period of the nonlinear
pulsation model are shown at the top of the plot.}
\label{figo54PW}
\end{figure}

\begin{figure}
\vspace{-3mm}
\includegraphics[scale=0.47]{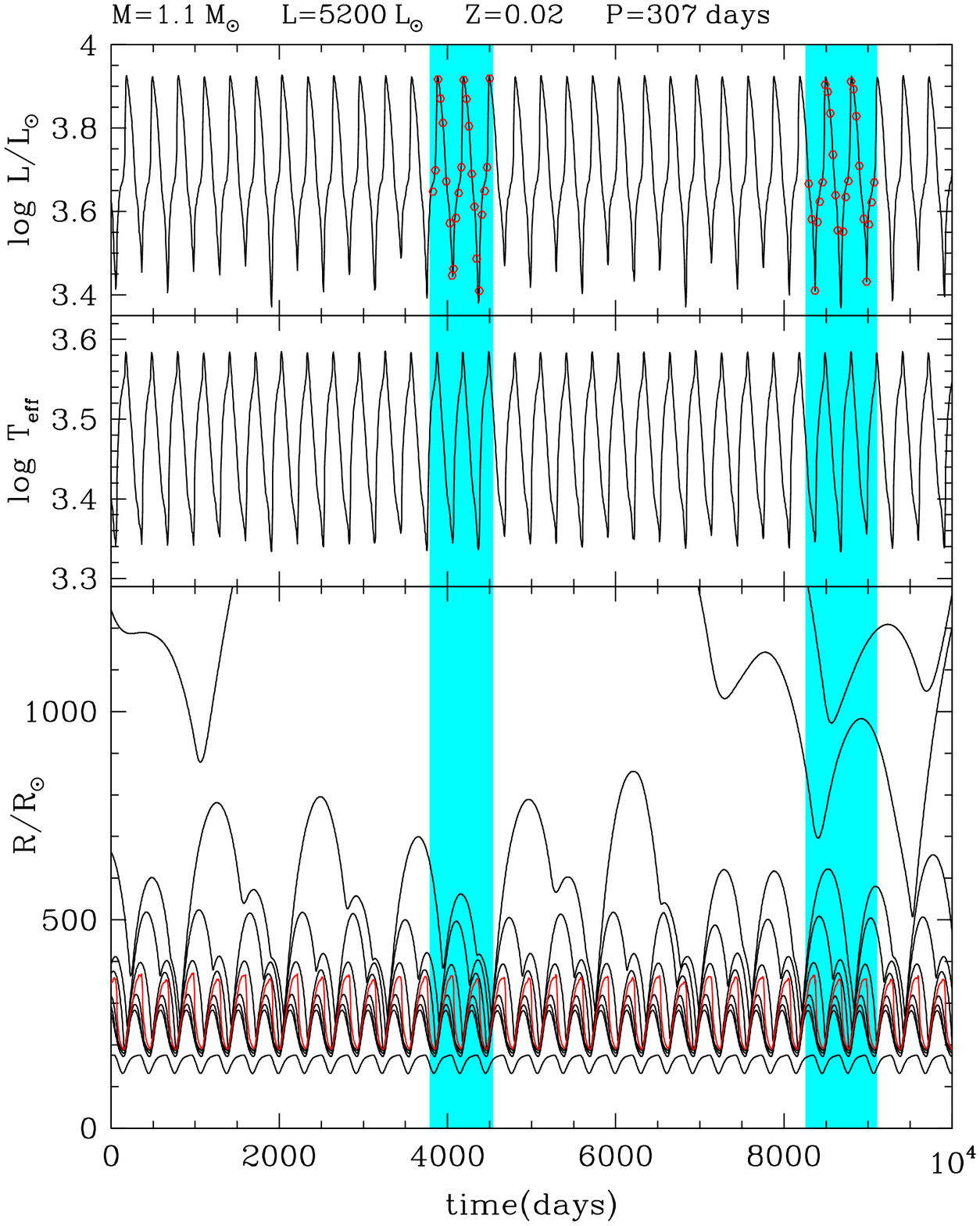}
\caption{The luminosity and mass-zone positions in the {\tt R52} series, showing the locations 
of the phases chosen for detailed model computation.}
\label{figR52PW}
\end{figure}

\begin{figure}
\vspace{-3mm}
\includegraphics[scale=0.47]{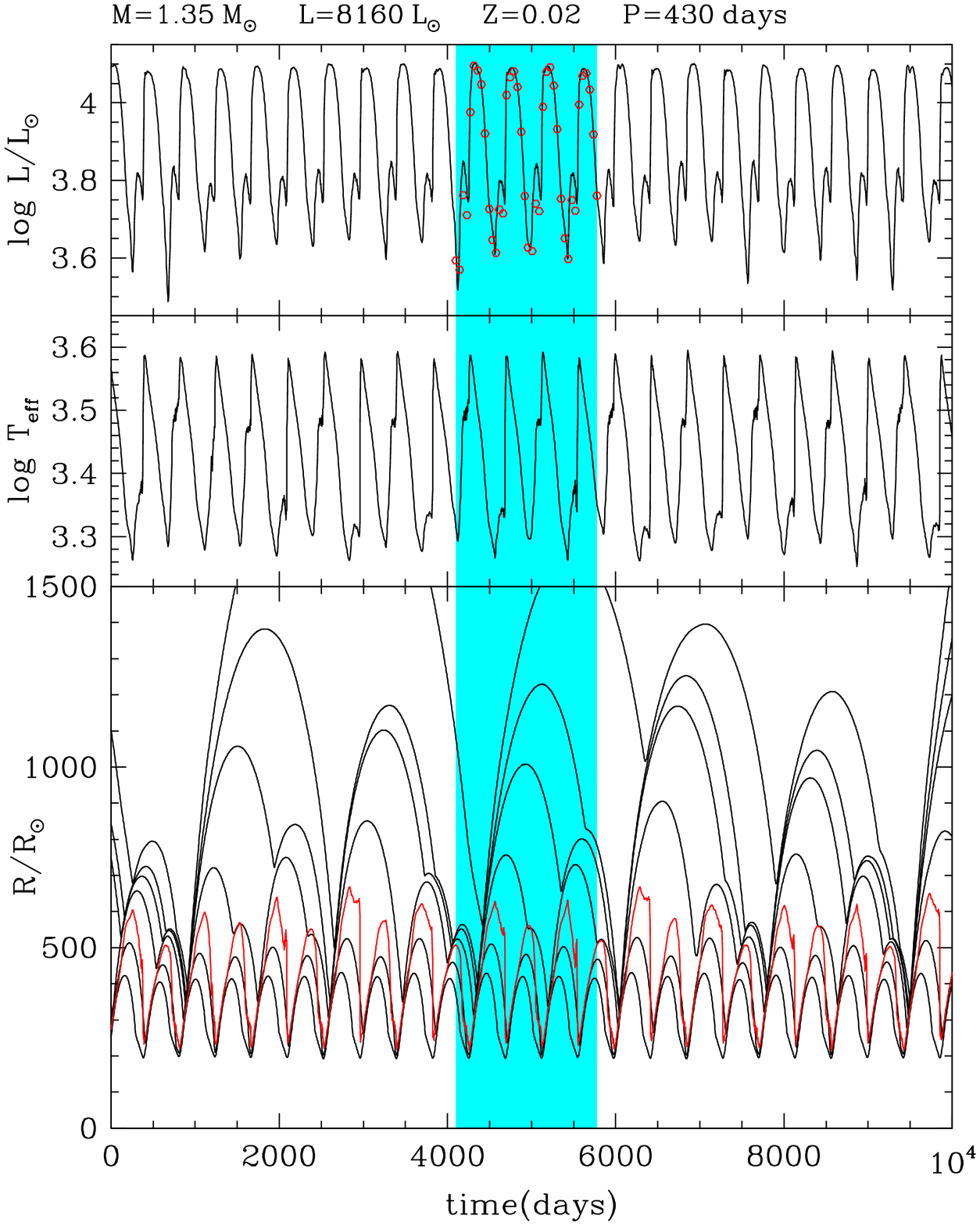}
\caption{The luminosity and mass-zone positions in the {\tt C81} series, showing the locations 
of the phases chosen for detailed model computation.}
\label{figC81PW}
\end{figure}

\begin{figure}
\vspace{-3mm}
\includegraphics[scale=0.47]{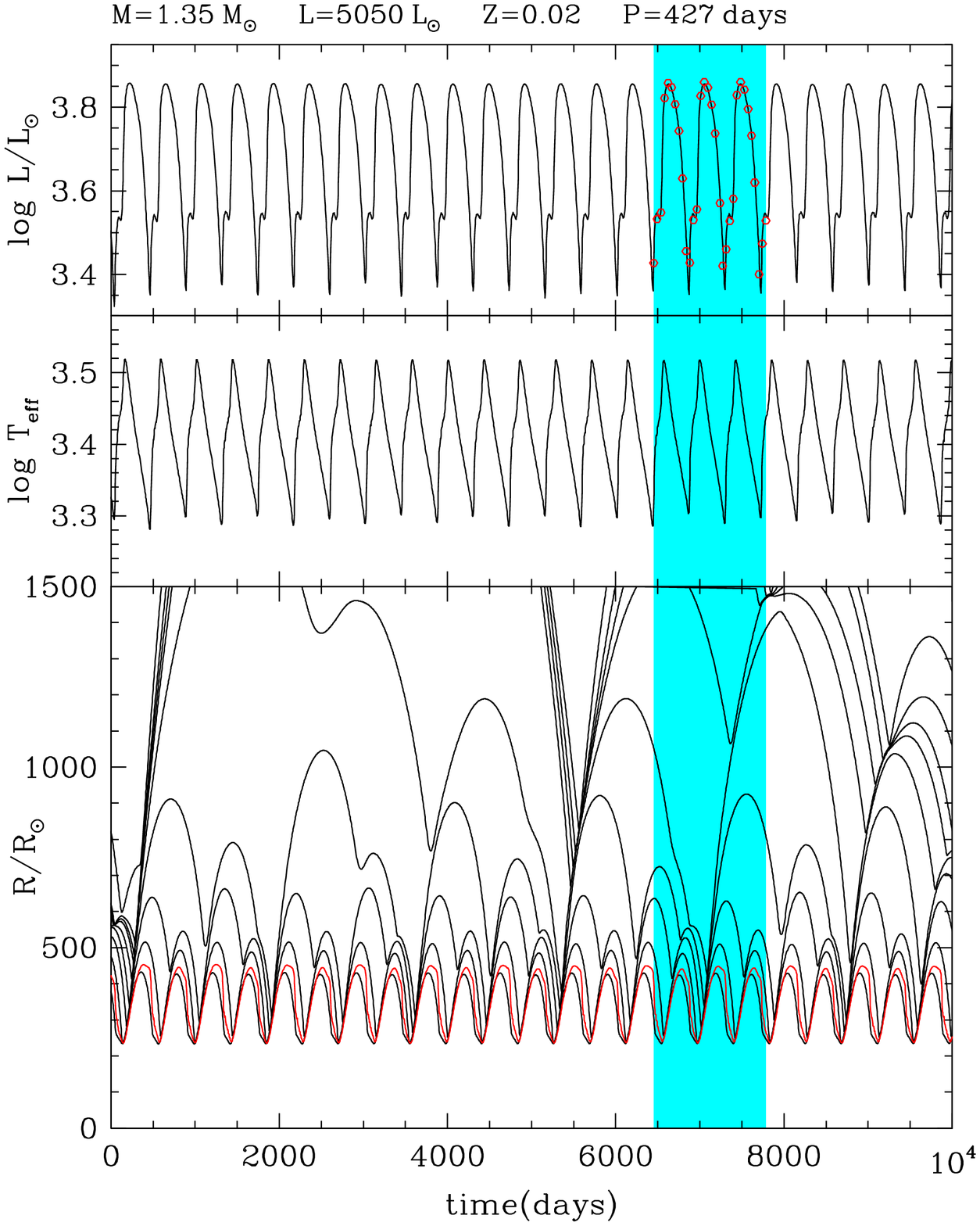}
\caption{The luminosity and mass-zone positions in the {\tt C50} series, showing the locations 
of the phases chosen for detailed model computation.}
\label{figC50PW}
\end{figure}

For all 4 model series, we assume solar element abundances from \citet{Grevesse96}. 
This is near the mean abundance
observed for stars in the solar vicinity with ages of 3--6$\times10^9$ years
\citep{Edvardsson93}.
Our red giant model with mass 1.35 M$_{\odot}$ has an age of
$\sim$3$\times10^9$ years and the red giant model with 
mass 1.1 M$_{\odot}$ has an age of
$\sim$6$\times10^9$ years according the isochrones of \citet{Girardi00}.
However, as shown by \citet{Edvardsson93}, there is a scatter 
in [Fe/H] from about -0.5 to +0.3 for stars of this age, so that
the Mira stars we are aiming to model could have somewhat different abundances
to those we have adopted.

The adopted value for the mixing length in units of pressure-scale
height for the {\tt o54}, {\tt R52} and
{\tt C81} series, $\alpha_m=3.5$, is unusually high compared with models of more
compact stars. However, this is not unreasonable, as detailed
hydrodynamic calculations of stellar convection often suggest
values for $\alpha_m$ in the
range 1--4 (see the parameter summary in \citealt{Meakin07}).
Decreasing the mixing length in models
makes heat transport more difficult in the outer convective layers, causing the
model star to expand. Therefore mixing length has been used by us as a
way to produce the correct period, given a luminosity $L$. However, this
procedure is always ambiguous for a given field Mira, because the
distance and hence $M$ is always uncertain by $\sim$10\% or more.

The value of $\alpha_{\nu}$ can be adjusted
to give the correct pulsation amplitude. Alternatively, within limits,
$M$ can adjusted to give the correct pulsation amplitude. Thus far in
our models series, we have assumed 
values of $M$ based on e.g. typical masses as a function of period and
then used $\alpha_{\nu}$ to tune the pulsation amplitude.
By comparing these model
series with observations, we aim to develop a preferred values for
$\alpha_m$ and $\alpha_\nu$, or at least a preferred prescription for
choosing $\alpha_m$ and $\alpha_\nu$ as a function of the physical input
parameters $M$, $L$ (a proxy for evolutionary state along the AGB) and $Z$. 

%

Although our models have been generally based on the stars $o$~Cet,
R~Leo and R~Cas, there are a range of parameters that are consistent
with these Miras, and a detailed comparison with observations will
inevitably reveal where the differences lie. In the following section, we aim to
discuss the chosen parameters for each model series,
compare predicted light curves to observed light curves and draw
preliminary conclusions as to whether changes to physical input 
parameters could improve model fits. We choose the
$V$, $J$ and $K$ band for comparison to observations: $V$ band because
of the wealth of observational data, and $J$ and $K$ because the model
predictions are most reliable in these band-passes. For this
comparison, we add 0.5 magnitudes to the predicted V-band fluxes, to
account for non-LTE effects as computed in the models at selected
phases in Paper~I, where the correction needed varies between 0.3 and
1.0 magnitudes. 

\begin{table}
\caption{Parameters of the extended {\tt o54} cycle, including the
  position of the shock fronts.}
\begin{tabular}{lrrrrlll}
Model & Phase & L& R &T$_{\rm eff}$ & S1 & S2 & S3 \\ 
      &       & ($L_\odot$)  & ($R_p$) & (K)    & ($R_p$) & ($R_p$) & ($R_p$) \\
\hline
260820&-0.20&  4050& 0.90& 3299&   2.82&  0.91 &       \\
260960&-0.10&  4240& 0.85& 3434&   2.85&  0.87 &       \\
261140& 0.00&  8013& 0.98& 3761&   2.86&  0.99 &       \\
261320& 0.10&  8102& 1.20& 3398&   2.87&  1.28 &       \\
261460& 0.19&  7420& 1.34& 3154&   2.83&  1.50 &       \\
261620& 0.31&  5830& 1.41& 2898&   2.74&  1.70 &       \\
261740& 0.40&  4462& 1.42& 2697&   2.64&  1.82 &       \\
261860& 0.50&  3640& 1.34& 2640&   2.47&  1.91 &       \\
261940& 0.60&  2440& 1.40& 2333&   2.31&  1.93 &       \\
262160& 0.70&  3450& 0.97& 3055&   2.07&  1.89  &  0.97  \\
262360& 0.80&  4110& 0.90& 3325&   $\Rightarrow$&  1.82&  0.90  \\
262600& 0.90&  4355& 0.85& 3462&       &  1.74&  0.86  \\
263160& 1.00&  8428& 0.99& 3786&       &  1.61&  1.01  \\
263740& 1.10&  8420& 1.20& 3439&       &  $\Rightarrow$&  1.37  \\
\end{tabular}
\label{tableo54_jx}
\end{table}

\begin{table}
\caption{Parameters of the compact {\tt o54} cycle, including the
  position of the shock fronts.[To be available online only]}
\begin{tabular}{lrrrrlll}
Model & Phase & L            & R &T$_{\rm eff}$ & S1 & S2 & S3 \\ 
      &       & ($L_\odot$)  & ($R_p$) & (K)    & ($R_p$) & ($R_p$) & ($R_p$) \\
\hline
248480&-0.30&  3243& 0.99& 2982&   2.03&  0.99 &       \\
248680&-0.20&  4122& 0.90& 3312&   2.02&  0.91 &       \\
248900&-0.10&  4379& 0.86& 3450&   1.99&  0.87 &       \\
249240& 0.00&  7870& 0.95& 3793&   1.92&  0.97 &       \\
249960& 0.10&  8358& 1.18& 3458&   1.80&  1.29 &       \\
250360& 0.21&  7432& 1.34& 3152&   $\Rightarrow$&  1.69 &       \\
250400& 0.29&  6239& 1.39& 2960&       &  2.10 &       \\
250420& 0.38&  4768& 1.41& 2749&       &  2.42 &       \\
250440& 0.53&  3520& 1.33& 2628&       &  2.96 &       \\
250460& 0.60&  2847& 1.32& 2497&       &  3.18&  1.13  \\
250640& 0.70&  3305& 0.99& 2990&       &  3.46&  0.99  \\
250820& 0.80&  4019& 0.91& 3287&       &  3.74&  0.91  \\
250980& 0.90&  4238& 0.85& 3436&       &  4.01&  0.86  \\
251160& 1.00&  7713& 0.96& 3768&       &  4.42&  0.97  \\
\end{tabular}
\label{tableo54_hx}
\end{table}

\begin{table*}
\caption{Parameters of the 4-cycle continuous phase coverage {\tt o54} cycle, including the
  position of the shock fronts.[To be available on-line only]}
\begin{tabular}{lrrrrllllll}
Model & Phase & L& R &T$_{\rm eff}$ & S1 & S2 & S3 & S4 & S5 & S6 \\ 
      &       & ($L_\odot$)  & ($R_p$) & (K)    & ($R_p$) & ($R_p$) & ($R_p$)& ($R_p$) & ($R_p$) & ($R_p$) \\
\hline
285180&-0.20&  3979& 0.92& 3263&   3.54&  0.92 &       &      &       &      \\
285320&-0.11&  4107& 0.86& 3397&   4.06&  0.87 &       &      &       &      \\
285500& 0.00&  7288& 0.93& 3769&   4.21&  0.94 &       &      &       &      \\
285700& 0.10&  8236& 1.17& 3468&   4.36&  1.22 &       &      &       &      \\
285860& 0.20&  7623& 1.33& 3189&   4.52&  1.47 &       &      &       &      \\
285980& 0.31&  6270& 1.40& 2952&       &  1.70 &       &      &       &      \\
286060& 0.41&  4513& 1.43& 2692&       &  1.87 &       &      &       &      \\
286100& 0.49&  3845& 1.40& 2614&       &  1.95 &       &      &       &      \\
286140& 0.59&  2548& 1.53& 2258&       &  2.02 &       &      &       &      \\
286320& 0.70&  3204& 1.00& 2959&       &  2.00 &  1.00 &      &       &      \\
286520& 0.80&  4092& 0.91& 3291&       &  1.93 &  0.92 &      &       &      \\
286700& 0.90&  4218& 0.86& 3415&       &  1.83 &  0.87 &      &       &      \\
287000& 1.00&  7188& 0.93& 3755&       &  1.64 &  0.94 &      &       &      \\
287560& 1.10&  8431& 1.17& 3487&       &  1.38 &  1.25 &      &       &      \\
287740& 1.20&  7734& 1.32& 3204&       &  $\Rightarrow$& 1.62 &      &       &      \\
287820& 1.30&  6304& 1.40& 2963&       &      &  1.95 &      &       &      \\
287880& 1.40&  4658& 1.43& 2720&       &      &  2.18 &      &       &      \\
287940& 1.51&  3820& 1.38& 2629&       &      &  2.38 &      &       &      \\
287980& 1.61&  2661& 1.47& 2326&       &      &  2.52 & 1.13 &       &      \\
288140& 1.70&  3342& 1.00& 2984&       &      &  2.61 & 1.00 &       &      \\
288320& 1.80&  4034& 0.91& 3279&       &      &  2.67 & 0.92 &       &      \\
288460& 1.90&  4159& 0.86& 3407&       &      &  2.71 & 0.87 &       &      \\
288620& 2.00&  7340& 0.93& 3770&       &      &  2.71 & 1.37 &       &      \\
288820& 2.10&  8170& 1.17& 3461&       &      &  2.71 & 1.22 &       &      \\
289020& 2.20&  7495& 1.32& 3181&       &      &  2.69 & 1.46 &       &      \\
289240& 2.30&  6174& 1.40& 2948&       &      &  2.66 & 1.66 &       &      \\
289440& 2.40&  4655& 1.42& 2730&       &      &  2.63 & 1.80 &       &      \\
289620& 2.49&  3918& 1.36& 2671&       &      &  2.46 & 1.91 &       &      \\
289740& 2.59&  2525& 1.42& 2336&       &      &  2.26 & 1.97 &       &      \\
289920& 2.70&  3064& 0.99& 2932&       &      &  $\Rightarrow$ & 1.99 & 0.99  &      \\
290120& 2.80&  4073& 0.91& 3290&       &      &       & 2.01 & 0.92  &      \\
290360& 2.90&  4301& 0.86& 3434&       &      &       & 2.08 & 0.87  &      \\
290740& 3.00&  7795& 0.95& 3796&       &      &       & 1.97 & 0.96  &      \\
291500& 3.10&  8412& 1.17& 3474&       &      &       & 2.06 & 1.29  &      \\
291740& 3.19&  7759& 1.32& 3210&       &      &       & $\Rightarrow$ & 1.65  &      \\
291800& 3.31&  6097& 1.40& 2935&       &      &       &      & 2.19  &      \\
291820& 3.41&  4451& 1.43& 2688&       &      &       &      & 2.55  &      \\
291840& 3.55&  3438& 1.40& 2544&       &      &       &      & 3.07  &      \\
291860& 3.61&  2813& 1.35& 2469&       &      &       &      & 3.24  &1.13  \\
\end{tabular}
\label{tableo54_fx}
\end{table*}

\begin{table*}
\caption{Parameters of the extended 2 cycles of the {\tt R52} series, including the
  position of the shock fronts.[To be available online only]}
\begin{tabular}{lrrrrllll}
Model & Phase & L& R &T$_{\rm eff}$ & S1 & S2 & S3 & S4  \\ 
      &       & ($L_\odot$)  & ($R_p$) & (K)    & ($R_p$) & ($R_p$) &
      ($R_p$)& ($R_p$) \\
\hline
360540&-0.20&  4442& 0.92& 3411&   2.02&  0.93 &       &      \\
360760&-0.10&  5000& 0.91& 3535&   1.94&  0.92 &       &      \\
361180& 0.00&  8266& 1.06& 3696&   1.79&  1.11 &       &      \\
361720& 0.10&  7425& 1.24& 3338&   1.57&  1.40 &       &      \\
361860& 0.19&  6491& 1.33& 3118&   $\Rightarrow$&  1.71 &       &      \\
361900& 0.32&  4700& 1.36& 2844&       &  2.10 &       &      \\
361920& 0.46&  3729& 1.25& 2793&       &  2.38 &       &      \\
361940& 0.54&  2795& 1.15& 2714&       &  2.51 &       &      \\
362020& 0.60&  2901& 1.06& 2851&       &  2.58&  1.06  &      \\
362200& 0.70&  3837& 0.96& 3209&       &  2.69&  0.97  &      \\
362380& 0.80&  4405& 0.92& 3400&       &  2.79&  0.94  &      \\
362560& 0.90&  5083& 0.91& 3545&       &  2.81&  0.92  &      \\
362780& 1.00&  8236& 1.06& 3704&       &  2.87&  1.10  &      \\
363000& 1.10&  7408& 1.24& 3337&       &  2.93&  1.37  &      \\
363180& 1.20&  6374& 1.33& 3097&       &  2.89&  1.59  &      \\
363380& 1.30&  4897& 1.36& 2870&       &  2.92&  1.77  &      \\
363540& 1.41&  4088& 1.31& 2797&       &  2.81&  1.91  &      \\
363600& 1.49&  3063& 1.24& 2674&       &  $\Rightarrow$&  1.97  &      \\
363700& 1.60&  2567& 1.06& 2767&       &      &  2.00 & 1.05  \\
363900& 1.70&  3910& 0.96& 3225&       &      &  2.00 & 0.97  \\
364120& 1.80&  4456& 0.92& 3409&       &      &  1.94 & 0.94  \\
364380& 1.90&  5076& 0.91& 3544&       &      &  1.86 & 0.93  \\
364960& 2.00&  8304& 1.07& 3699&       &      &  1.76 & 1.12  \\
\end{tabular}
\label{tableR52_gx}
\end{table*}

\begin{table*}
\caption{Parameters of the compact 2 cycles of the {\tt R52} series, including the
  position of the shock fronts.[To be available on-line only]}
\begin{tabular}{lrrrrllll}
Model & Phase & L& R &T$_{\rm eff}$ & S1 & S2 & S3 & S4  \\ 
      &       & ($L_\odot$)  & ($R_p$) & (K)    & ($R_p$) & ($R_p$) &
      ($R_p$)& ($R_p$) \\
\hline
386260& 2.36&  4640& 1.35& 2836&   2.18&      &       &      \\
386280& 2.49&  3813& 1.26& 2796&   2.56&      &       &      \\
386320& 2.60&  2569& 1.13& 2680&   2.89&  1.10&       &      \\
386500& 2.70&  3753& 0.99& 3148&   3.09&  1.00&       &      \\
386660& 2.80&  4198& 0.93& 3332&   3.27&  0.95&       &      \\
386840& 2.90&  4680& 0.90& 3486&   3.46&  0.92&       &      \\
387040& 3.00&  8026& 0.99& 3801&       &  1.02&       &      \\
387260& 3.10&  7704& 1.18& 3446&       &  1.27&       &      \\
387420& 3.19&  6844& 1.30& 3193&       &  1.49&       &      \\
387580& 3.30&  5449& 1.36& 2951&       &  1.70&       &      \\
387700& 3.40&  4350& 1.34& 2805&       &  1.85&       &      \\
387780& 3.49&  3585& 1.27& 2743&       &  1.93&       &      \\
387860& 3.61&  2049& 1.24& 2421&       &  1.98&  1.09 &      \\
388040& 3.70&  3560& 0.99& 3111&       &  1.97&  0.99 &      \\
388240& 3.80&  4311& 0.93& 3355&       &  1.92&  0.95 &      \\
388480& 3.90&  4704& 0.90& 3483&       &  1.83&  0.93 &      \\
388900& 4.00&  8166& 0.99& 3814&       &  1.69&  1.03 &      \\
389540& 4.10&  7818& 1.17& 3473&       &  1.48&  1.30 &      \\
389680& 4.20&  6744& 1.31& 3172&       &  $\Rightarrow$&  1.69 &      \\
389720& 4.32&  5120& 1.35& 2908&       &      &  2.06 &      \\
389740& 4.49&  3819& 1.26& 2803&       &      &  2.45 &      \\
389780& 4.60&  2700& 1.12& 2728&       &      &  2.65 & 1.10 \\
389940& 4.69&  3700& 0.99& 3128&       &      &  2.77 & 1.00 \\
390120& 4.80&  4182& 0.93& 3332&       &      &  2.88 & 0.95 \\
390300& 4.90&  4674& 0.90& 3482&       &      &  3.01 & 0.93 \\
\end{tabular}
\label{tableR52_fx}
\end{table*}

\begin{table*}
\caption{Parameters of the {\tt C50} series, including the
  position of the shock fronts.[To be available on-line only]}
\begin{tabular}{lrrrrlllll}
Model & Phase & L& R &T$_{\rm eff}$ & S1 & S2 & S3 & S4 & S5 \\ 
      &       & ($L_\odot$)  & ($R_p$) & (K)    & ($R_p$) & ($R_p$) &
      ($R_p$)& ($R_p$) & ($R_p$)\\
\hline
375360&-1.30&  2673& 0.92& 2532&   2.49&  0.92&       &      &       \\
375490&-1.20&  3407& 0.85& 2810&   2.56&  0.87&       &      &       \\
375630&-1.10&  3529& 0.80& 2924&   2.59&  0.83&       &      &       \\
375780&-1.00&  6640& 0.87& 3271&   2.63&  0.92&       &      &       \\
375920&-0.90&  7218& 1.03& 3077&   2.62&  1.13&       &      &       \\
376050&-0.80&  7039& 1.16& 2883&   2.60&  1.37&       &      &       \\
376160&-0.70&  6412& 1.23& 2738&   2.55&  1.57&       &      &       \\
376260&-0.60&  5537& 1.24& 2627&   2.48&  1.74&       &      &       \\
376330&-0.50&  4263& 1.21& 2493&   2.38&  1.84&       &      &       \\
376380&-0.40&  2851& 1.10& 2357&   2.19&  1.91&       &      &       \\
376470&-0.30&  2678& 0.92& 2543&   $\Rightarrow$&  1.93& 0.92&      &       \\
376630&-0.20&  3392& 0.84& 2816&       &  1.97&  0.87 &      &       \\
376820&-0.10&  3595& 0.80& 2937&       &  1.97&  0.82 &      &       \\
377110& 0.00&  6722& 0.88& 3261&       &  1.86&  0.94 &      &       \\
377490& 0.10&  7253& 1.04& 3058&       &  1.75&  1.18 &      &       \\
377710& 0.20&  7037& 1.16& 2876&       &  $\Rightarrow$&  1.53 &      &       \\
377750& 0.30&  6400& 1.23& 2735&       &      &  2.01 &      &       \\
377760& 0.40&  5465& 1.24& 2618&       &      &  2.39 &      &       \\
377770& 0.53&  3720& 1.21& 2407&       &      &  2.83 &      &       \\
377790& 0.61&  2633& 1.18& 2233&       &      &  3.06 &      &       \\
377880& 0.70&  2883& 0.91& 2606&       &      &  3.31 & 0.91 &       \\
378020& 0.80&  3370& 0.84& 2823&       &      &  3.55 & 0.86 &       \\
378160& 0.90&  3813& 0.80& 2981&       &      &  3.76 & 0.82 &       \\
378320& 1.00&  6750& 0.89& 3253&       &      &  3.96 & 0.93 &       \\
378470& 1.10&  7246& 1.06& 3035&       &      &  4.15 & 1.18 &       \\
378570& 1.21&  6955& 1.18& 2852&       &      &  4.34 & 1.44 &       \\
378630& 1.31&  6250& 1.23& 2713&       &      &  4.52 & 1.65 &       \\
378660& 1.40&  5395& 1.24& 2610&       &      & (4.66)& 1.80 &       \\
378680& 1.49&  4166& 1.21& 2476&       &      & (4.80)& 1.91 &       \\
378710& 1.61&  2512& 1.06& 2326&       &      & (4.97)& 1.97 &       \\
378810& 1.70&  2972& 0.90& 2631&       &      &       & 1.95 & 0.91  \\
378960& 1.80&  3380& 0.83& 2827&       &      &       & 1.89 & 0.86  \\
\end{tabular}
\label{tabler50}
\end{table*}

\begin{table*}
\caption{Parameters of the {\tt C81} series, including the
  position of the shock fronts.[To be available on-line only]}
\begin{tabular}{lrrrrllllll}
Model & Phase & L& R &T$_{\rm eff}$ & S1 & S2 & S3 & S4 & S5 & S6 \\ 
      &       & ($L_\odot$)  & ($R_p$) & (K)    & ($R_p$) & ($R_p$) & ($R_p$)& ($R_p$) & ($R_p$) & ($R_p$) \\
\hline
243570&-0.40&  3917& 1.58& 2182&   2.74&  1.96&       &      &       \\
243670&-0.30&  3705& 1.34& 2336&   2.40&  2.06&  0.98 &      &       \\
244000&-0.20&  5772& 0.87& 3235&   $\Rightarrow$&  2.14&  0.87 &      &       \\
244140&-0.10&  5133& 0.79& 3311&       &  2.29&  0.79 &      &       \\
244330& 0.00&  9462& 0.81& 3811&       &  2.25&  0.81 &      &       \\
244700& 0.10& 12460& 1.08& 3522&       &  2.21&  1.14 &      &       \\
246230& 0.20& 12120& 1.28& 3212&       &  2.20&  1.49 &      &       \\
246810& 0.30& 11150& 1.40& 3015&       &  2.05&  1.87 &      &       \\
246910& 0.40&  8327& 1.44& 2758&       &  $\Rightarrow$& 2.31&      &       \\
246940& 0.52&  5324& 1.65& 2310&       &      &  2.82 &      &       \\
246960& 0.61&  4427& 1.84& 2087&       &      &  3.07 &      &       \\
247020& 0.70&  4103& 1.63& 2175&       &      &  3.39 & 1.01 &       \\
247280& 0.80&  5302& 0.89& 3130&       &      &  3.67 & 0.89 &       \\
247450& 0.90&  5184& 0.79& 3322&       &      &  3.88 & 0.79 &       \\
247560& 1.00& 10470&     &     &       &      &  4.13 & 0.78 &       \\
247690& 1.10& 11640& 1.06& 3498&       &      &  4.39 & 1.08 &       \\
247810& 1.20& 12060& 1.27& 3228&       &      &  4.54 & 1.35 &       \\
247900& 1.30& 10990& 1.39& 3014&       &      & (4.80)& 1.60 &       \\
247970& 1.40&  8416& 1.44& 2769&       &      & (4.92)& 1.79 &       \\
248020& 1.50&  5748& 1.58& 2405&       &      &       & 1.91 &       \\
248060& 1.59&  4234& 1.67& 2169&       &      &       & 1.99 &       \\
248130& 1.70&  4143& 1.41& 2341&       &      &       & 2.00 & 1.00  \\
248430& 1.80&  5493& 0.88& 3175&       &      &       & 1.96 & 0.88  \\
248580& 1.90&  5250& 0.79& 3326&       &      &       & 1.89 & 0.79  \\
248730& 2.00&  9757& 0.79& 3882&       &      &       & 1.76 & 0.79  \\
249030& 2.10& 12020& 1.06& 3523&       &      &       & 1.59 & 1.09  \\
249400& 2.20& 12350& 1.26& 3255&       &      &       & $\Rightarrow$ & 1.41  \\
249500& 2.30& 11070& 1.38& 3029&       &      &       &      & 1.82  \\
249580& 2.39&  8557& 1.42& 2801&       &      &       &      & 2.09  \\
249710& 2.50&  5655& 1.57& 2400&       &      &       &      & 2.35  \\
249890& 2.60&  4468& 1.68& 2191&       &      &       &      & 2.53  \\
250020& 2.70&  3955& 1.59& 2180&       &      &       &      & 2.68  \\
250290& 2.80&  5609& 0.89& 3178&       &      &       &      & 2.79  \\
250440& 2.89&  5270& 0.80& 3311&       &      &       &      & 2.88  \\
250550& 3.00&  9899&     &     &       &      &       &      & 2.94  \\
250670& 3.10& 11740& 1.06& 3501&       &      &       &      & 2.97  \\
250780& 3.20& 11920& 1.27& 3217&       &      &       &      & 2.98  \\
250870& 3.30& 10810& 1.38& 3012&       &      &       &      & 2.98  \\
250960& 3.40&  8299& 1.43& 2765&       &      &       &      & 2.95  \\
251040& 3.49&  5758& 1.55& 2429&       &      &       &      & 2.87  \\
\end{tabular}
\label{tabler81}
\end{table*}

\subsection{{\tt o54} 5400\,$L_{\odot}$ series}

The 5400\,L$_{\odot}$ model series for $o$~Cet ($P = 332$\,days) has model parameters
chosen to match the  luminosity obtained from the $J$ and $K$
photometry of \citet{Whitelock00} and the \citet{Knapp03} revised {\tt HIPPARCOS} parallax.
The mass of 1.1\,M$_\odot$ was adopted since an analysis of the
Galactic scale height
of Mira variables by \citet{Jura92}
suggest a mass of $\sim$1.1\,M$_\odot$ for Mira variables with periods from 300--400\,days,
while a study of the population of Miras by Wyatt and Cahn (1983)
gives a progenitor mass estimate of 1.18\,M$_\odot$ for $o$~Cet.
We give this model series the designation {\tt o54}. 
 
\begin{figure}
\includegraphics[scale=0.7]{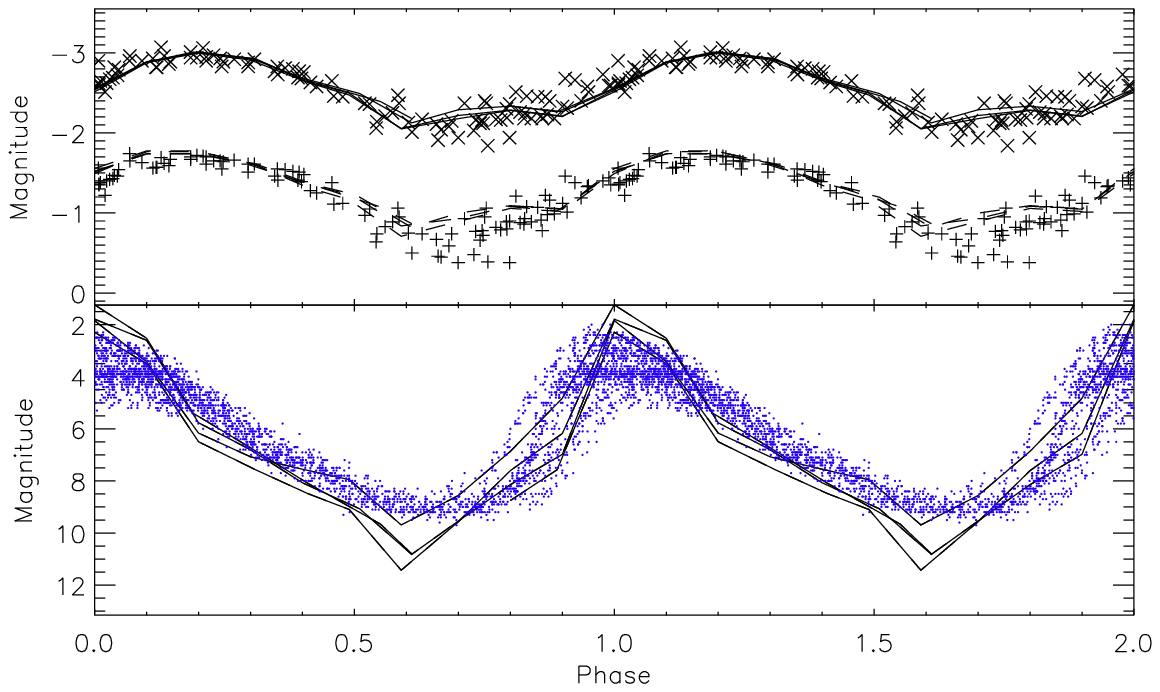}
\caption{Light curves of the {\tt o54} series in $J$, $K$ and $V$ bands
  compared with observations of $o$~Cet by \citet{Whitelock00} and the AFOEV.}
\label{figoCetLCurve}
\end{figure}

Figure~\ref{figoCetLCurve} shows the light curves derived from 4
cycles of the {\tt o54} model series between times 8000 
and $10^4$ days in Figure\,\ref{figo54PW}.  To fit the K-band maximum flux,  $o$~Cet
needs to be at a distance of
107\,pc . This corresponds to a parallax
of 9.3\,mas, consistent within 2$\sigma$ of the latest 
HIPPARCOS value of 10.91$\pm$1.22\,mas from \citet{vanLeeuwen07}.

The models fit the general light-curve shape and amplitude, but they are
$\sim$0.1\,magnitudes too blue in $J$-$K$ colour, 
too blue in $V$-$K$ by $\sim$1 magnitudes (hence too bright in $V$ near maximum
by $\sim$1 magnitudes), and these 4 cycles do not reproduce the 
cycle-to-cycle
scatter in $J$ and $K$ magnitudes near minimum. 

We also computed models for more compact atmosphere cycles 
(between times 1000 and 2000 days in Figure 1) and more
extended atmosphere cycles (between times 3500 and 4300 days), but $J$, $H$ and $V$ fluxes
were not noticeably different. The atmospheric structures for the various
models are shown in Figure~\ref{figo54PW}. The consistently blue $J$-$K$ and $V$-$K$
colours suggest that Teff is too high in the model (see Section 5 for
more discussion).

 Although the basic
model properties ($L$, $R$, $T_{\rm eff}$) and the near-continuum
fluxes are similar from cycle to cycle, high-layer observational
features differ depending on the upper atmosphere structure. The
reason for this can be seen in Tables~\ref{tableo54_jx} through \ref{tableo54_fx}, where
upper layer shock front positions are not repeatable from cycle to
cycle. For example, in Table~\ref{tableo54_jx}, at phase -0.2, the
upper strong shock from a previous cycle is at 2.82\,$R_p$. Exactly 1
cycle later, at phase 0.8, the upper shock has just merged with a
lower shock at 1.82\,$R_p$.

\subsection{{\tt R52} 5200\,$L_{\odot}$ series}

The period of R~Leo is slightly shorter than that of $o$~Cet, 
and we chose to model
R~Leo with a model identical to the $o$~Cet model, but with a luminosity reduced
to match the smaller period (310 versus 332 days).  Miras typically
also have reduced masses at reduced periods \citep{Jura92}, but we
chose not to reduce the mass so as to see the differential influence
of luminosity alone (Wyatt and Cahn 1983 give a mass of 1.04\,M$_{\odot}$
for R~Leo).  The
model has a luminosity derived by assuming R~Leo was at a distance of 110\,pc, corresponding to a
parallax of 9.1\,mas, again consistent with the \citet{vanLeeuwen07}
value of 14.03$\pm$2.65\,mas within 2-$\sigma$.

\begin{figure}
\includegraphics[scale=0.7]{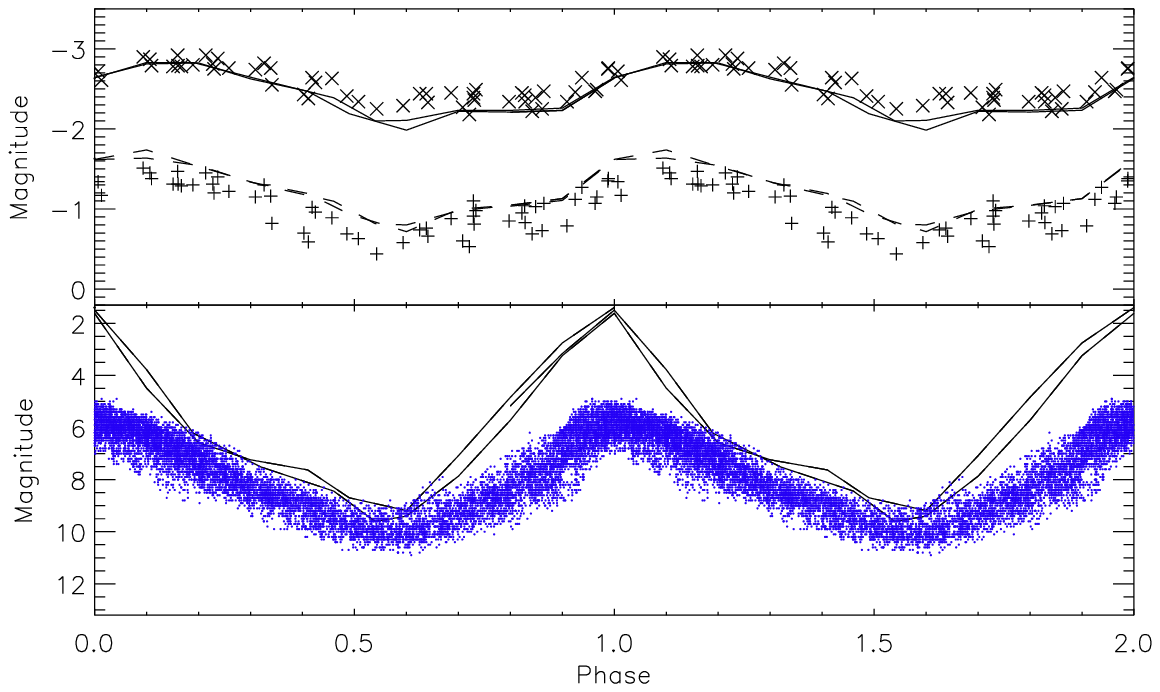}
\caption{Light curves of the {\tt R52} series in the extended cycle
  in $J$, $K$ and $V$ bands compared with
  observations by \citet{Whitelock00} and the AFOEV.}
\label{figRLeoLCurve}
\end{figure}

It is clear in Figure~\ref{figRLeoLCurve} that
this model is too blue to be an effective model for R~Leo in both $V$-$K$ and $J$-$K$ colours, and has a visual
amplitude much larger and a $K$ amplitude slightly larger than R~Leo.  
The amplitude of the R~Leo model is slightly smaller
than the amplitude of the $o$~Cet model due to the reduced luminosity.
This luminosity change by itself is
not enough to explain the different visual amplitudes of the real R~Leo and
$o$~Cet.  It is possible that these two stars have different metallicity
(i.e. the redder colours of R Leo could be because it has a higher metallicity).
A difference in mass is also possible, with a compensating change
in luminosity within that allowed by the parallax error to retain the same period. 

Like the {\tt o54} series, we computed an extended sub-series (day
numbers 3600--4500 in Figure~\ref{figR52PW}) and a compact sub-series
(day numbers 8200--9100 in Figure~\ref{figR52PW}). The compact
sub-series had its detailed radiative transfer model truncated at
4\,$R_p$, because the very low density in the outer layers
($<10^{-15}$\,g\,cm$^{-3}$) had extremely low opacity and the models
had outer-layer physical conditions outside the range where our
equation of state was valid.
Again, the $V$, $J$ and $K$ fluxes were similar in each case (only the
extended sub-series is displayed). 

\subsection{{\tt C81} 8160\,$L_{\odot}$ series}
With the same input physics as the $o$~Cet models, we attempted to
create a longer period series appropriate for the Mira variable R~Cas ($P = 430$\,days)
by increasing the luminosity and
mass. However, the model pulsation amplitude became much too large,
requiring us to increase the $\alpha_\nu$ parameter. The
HIPPARCOS distance for R~Cas in \citet{Whitelock08} would require the star to be very
under-luminous when compared with the mean solar-vicinity P-L relationship of \citet{Whitelock08}
or the LMC P-L relationship \citep[e.g][]{Hughes90}.  For this model series, the luminosity of our model
was derived by assuming R~Cas falls on the mean solar-vicinity P-L relationship of \citet{Whitelock08} and 
is at a distance of 204\,pc, with a corresponding parallax of
4.9\,mas, almost 3-$\sigma$ from the \citet{vanLeeuwen07} value of
7.95$\pm$1.02\,mas.   The mass adopted for R~Cas is 1.35 M$_{\odot}$ as
given by \citet{Wyatt83}.

The light curves of R~Cas in the $V$, $J$ and $K$ bands are shown in
Figure~\ref{figRCas8Curve}, corresponding to day numbers $\sim$4000-5700 chosen for
detailed radiative transfer computation in Figure~\ref{figC81PW}.
The $J$ and $K$ light curves of the model fit the observations quite well,
while in all cycles where the near-maximum model was computed, the
models are too bright in $V$ near maximum light. 
The near-maximum continuum effective temperature
is $\sim$3800\,K in Table~\ref{tabler81}, corresponds to an ~M0 or
M1 giant according to the temperature calibration of \citet{Fluks94} and appears like
an M2 giant in the TiO features as predicted by our model
spectra. This is much too warm for R~Cas which has
a catalogued spectral type of M6--M10. Therefore, we are forced to conclude that
$T_{\rm eff}$ is too high for this model. 

\subsection{{\tt C50} 5050\,$L_\odot$ series}
As the 8160\,L$_{\odot}$ model or R~Cas was so clearly discrepant near
maximum, and as the individual HIPPARCOS distance in Whitelock et al. (2008)
would give a luminosity of only $\sim$3770\,L$_{\odot}$ ,
we chose to construct a lower luminosity model.  The
luminosity was determined by fixing the mixing length parameter
which was decreased to a more
standard value of $\alpha_m=2$.  The luminosity was then tuned to match the
model period to that of R~Cas.  As usual, the turbulent viscosity parameter 
was then tuned to match the
bolometric amplitude of R~Cas and the model. The resulting luminosity (5050\,L$_{\odot}$)
suggests that R~Cas is at
a distance of 166\,pc with a corresponding parallax of 6.0\,mas: this is now
consistent with the HIPPARCOS value within 2-$\sigma$.

\begin{figure}
\includegraphics[scale=0.7]{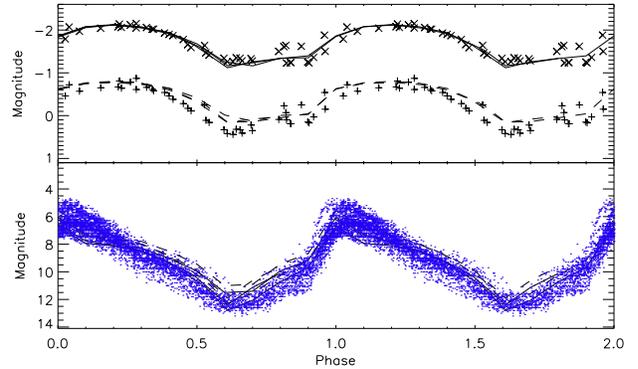}
\caption{Light curves of the {\tt C50} series $J$, $K$ and $V$ bands compared with
  observations by \citet{Nadzhip01} and the AFOEV.}
\label{figRCas5LCurve}
\end{figure}

\begin{figure}
\includegraphics[scale=0.7]{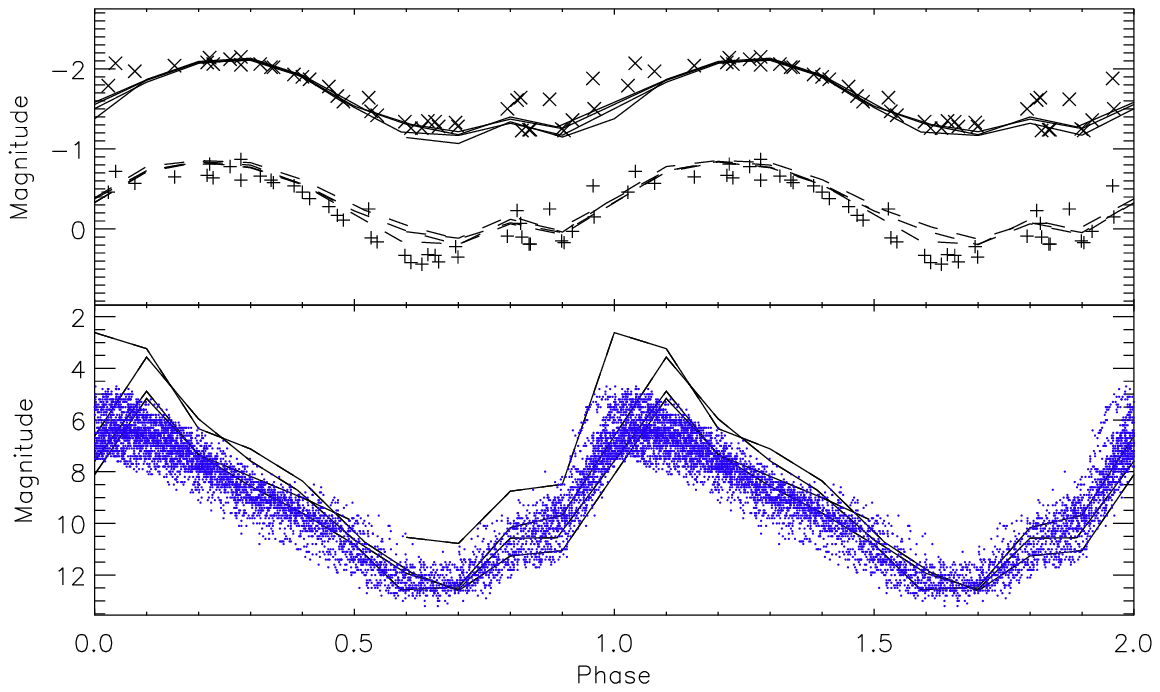}
\caption{Light curves of the {\tt C81} series $J$, $K$ and $V$ bands compared with
  observations by \citet{Nadzhip01} and the AFOEV.}
\label{figRCas8Curve}
\end{figure}

The near-maximum effective temperature of this model is now
$\sim3250$\,K, both consistent with an M6 spectral type from the calibration
of \citet{Fluks94} and providing a better match to the observed M6
spectra in \citet{Fluks94} than either M5 of M7 spectra. It also has a
a $V$-$K$ colour that matches that of R~Cas (see Figure~\ref{figRCas5LCurve}).

\subsection{Using the Models}
\label{sectUsingModels}

\begin{table*}
\caption{An excerpt of the table available on-line describing the model
  output for the {\tt o54} series model 285180. The center-to-limb
  variation (CLV) is shown as a fraction of the central intensity
  $I_0$ as a function of normalized radius on the apparent stellar
  disk, r/5R$_p$.}
\begin{tabular}{lllrrrrrrrrrr}
Wavelength & L$_\lambda$ & I$_0$  & & CLV:  &  r/5R$_p$& and &I/I$_0$&  & & & \\
($\mu$m)   & (erg s $\mu$m$^{-1}$) & (erg s cm$^{-2}$ $\mu$m$^{-1}$ sr$^{-1}$) & 
0.020 & 0.100 & 0.140 & 0.170 & 0.190 & 0.220 & 0.290 & 0.360 & 0.440 \\
\hline
1.598  & 0.9786E+37 & 0.1124E+10 & 0.997 & 0.923 & 0.828 & 0.691 & 0.451 & 0.082 & 0.066 & 0.053 & 0.046 \\
1.599  & 0.8160E+37 & 0.3054E+09 & 0.999 & 0.977 & 0.955 & 0.932 & 0.913 & 0.879 & 0.796 & 0.662 & 0.337\\
1.600  & 0.1106E+38 & 0.1427E+10 & 0.998 & 0.940 & 0.863 & 0.743 & 0.274 & 0.024 & 0.022 & 0.022 & 0.026\\
\end{tabular}
\label{tabModelOutput}
\end{table*}

For each model phase in Tables~\ref{tableo54_hx} through
\ref{tabler81}, we provide the full model output
\footnote{http://www.physics.mq.edu.au/$\sim$mireland/codex/}. 
Indeed, this output was already used by \citet{Woodruff09} in advance
of publication in order to compare the models to wavelength-dispersed
infrared interferometry. A sample
3 lines from one of these tables is given in
Table~\ref{tabModelOutput}. In order to use these models to compare to
a specific observation, integration over a filter profile
$F(\lambda)$ is required:

\begin{equation}
 L_F = \int_\lambda L(\lambda) d\lambda
\end{equation}

\begin{equation}
 I_F(x) = \int_\lambda F(\lambda) I_0(\lambda) f(x,\lambda) d\lambda.
\end{equation}

Here $L_F$ is the stellar luminosity as seen through the filter,
$I_F(x)$ is the intensity profile seen through the filter, $I_0(\lambda)$ are
the tabulated values of the central intensity and $f(x,\lambda)$ are
the tabulated values of the normalized CLVs. It is much more preferable to use realistic filter 
profiles $F(\lambda)$ 
with smooth edges (e.g. a Gaussian) rather than square-edged filters in order to minimise noise
due to the opacity sampling.
Interferometric visibilities can
then be obtained from the Hankel transform of $I_F(x)$. 

\section{Model Spectra}
\label{sectSpectra}

The spectra computed in the {\small CODEX} models using the default
wavelength table come from an opacity sampling method with a spectral
resolution of up to $\sim$10$^4$. However,  in order to accurately
compare with observations at any wavelength, at least $\sim$100
wavelengths have to be averaged together, preferably using a
non-square edged filter. In turn, this means that these default model
outputs can only be used at a spectral resolution of $R\sim100$ or
lower. This is especially true where the CO bands in $H$ and $K$ bands are
concerned, where there is a combination of very strong absorption and
near-continuum spectral features. 

\begin{figure}
  \includegraphics[scale=0.6]{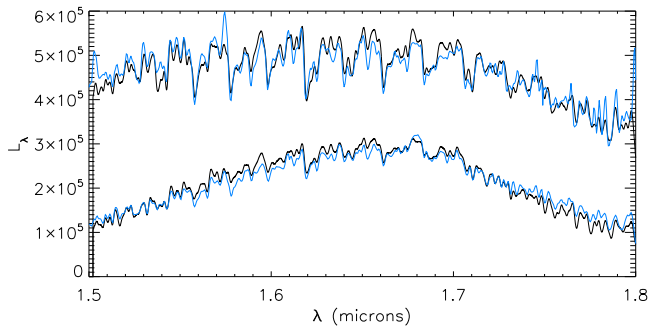}
  \caption{A H-band comparison of the R~Cha spectrum at phases 0.1
  (top) and 0.3 (bottom) to the
  287820 (top) and 250460 (bottom) models of the {\tt o54} series.}
  \label{figSpectraHComp}
\end{figure}

\begin{figure}
  \includegraphics[scale=0.6]{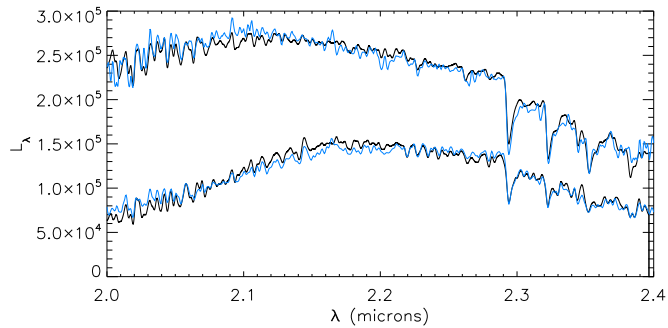}
  \caption{A K-band comparison of the R~Cha spectrum at phases 0.1
  (top) and 0.3 (bottom) to the
  287820 (top) and 250460 (bottom) models of the {\tt o54} series.}
  \label{figSpectraKComp}
\end{figure}

For the purposes of comparing spectra with observations in H- and
K-bands, we therefore also computed spectra and center-to-limb
variations at R$\approx10^5$ in these bands. In
Figures~\ref{figSpectraHComp} and \ref{figSpectraKComp} we convolved
the model spectra with a 
Gaussian of Full-Width Half Maximum equivalent to a spectral
resolution of $R=1000$, and compared the model spectra with
observations of the Mira variable R~Cha from \citet{Lancon00}. We
chose to compare with the 
{\tt o54} series, because the parameters of R~Cha are most like those
of $o$~Cet. There was an arbitrary scaling applied to the observed
spectra. These factors differed by 0.05 mags between the H- and K-
bands for the phase 
0.1 spectrum and 0.15 magnitudes for the phase 0.3 spectrum: these small
differences . The best model
fits were from phases 0.3 and 0.6, where temperatures were
$\sim$500\,K cooler than phases applicable to the R~Cha
observations. This demonstrates that in the CO overtone bands, the
models are too hot for R~Cha. To make this statement more quantitatively based on spectral
synthesis, metallicity effects would have to be considered also
(beyond the scope of this paper).

%

\section{Observation and Model Comparison: $o$~Cet}
\label{sectInterferometry}

The grand total of all available observations for Mira variables with
parameters similar to those of the model series presented here is far
too vast to compare to the model series of this paper in a concise
manner. Therefore, we have chosen to examine the available
observations of $o$~Cet in a general sense to describe the
similarities and differences between the {\tt o54} series and
$o$~Cet to further illustrate the utility of the model series and the wealth 
of information available to constrain models.

Table~\ref{tbloCetObs} summarizes most of the key observations
available for $o$~Cet. Time-dependent photometry is available between
ultraviolet and radio wavelengths, with the best light curves
available in the $V$, $J$, $H$ and $K$ bands, as shown in
Figure~\ref{figoCetLCurve} (V,J,K). There is reasonable agreement between 
the model and observations for light curve shape, amplitude and visible-infrared phase offsets.

Spectral classification should also give an observed effective temperature. Unfortunately, the 
MK spectral classification \citep[e.g][]{Keenan74} is based on B and $V$ bands, where non-LTE effects 
in an extended atmosphere are very strong \citep{Ireland08}. Consequently, the combination of
effective temperature and metallicity can not be directly fit to observations. Spectra of $o$~Cet are also 
not available electronically to the knowledge of the authors -- a modern library of bright Mira spectra
would certainly be of great use to future modelling efforts. In particular, infrared spectra are a much more
reliable model output, and phase-dependent infrared spectra would be a wonderful tool for tuning model parameters.

\begin{figure}
\includegraphics[scale=0.7]{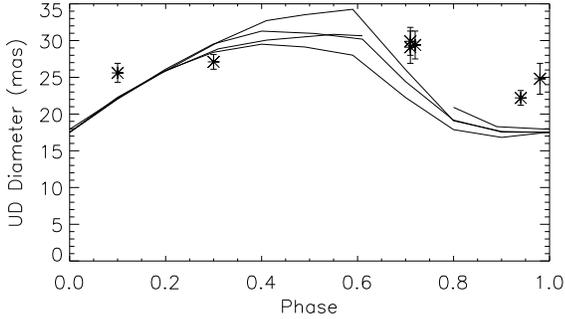} 
\caption{Model diameters of the {\tt o54} series in a narrow 1.24\,$\mu$m bandpass based on
  fitting to a single spatial frequency where $V=0.6$, with the
  measured diameters from \citet{Woodruff08} over-plotted.}  
\label{figJDiams}
\end{figure}

Resolved observations have been made at
wavelengths between 346\,nm and 7\,mm, with the broad range of highly
wavelength dependent diameters being consistent with Models. The form
of angular diameter versus wavelength curve as shown in
\citet{Woodruff09} between 1 and 4 microns was very similar to the {\tt o54} model series,
however, in that paper, the models were placed at a distance that best
fit the angular diameters. If instead the models are placed at the
distance that best fits the K-band photometry, the angular diameters
as a function of phase are given in Figures~\ref{figJDiams}
to~\ref{figKDiams}. It is clear that the mean diameter of the models
are too small, and that the phase-dependence of 
the observed diameter is less pronounced than in the model. 
Possible solutions to this are given in
Sections~\ref{sectParams} and~\ref{sectRadAccel}.

 In the 
radio, the measured angular size of $o$~Cet corresponds to the
angular size of the Na and/or $K$ ionisation edges. The consistency
between models and observations shows that local thermal equilibrium
is a reasonably assumption for defining the ionization fraction. 
In the ultraviolet, the measured FWHM of 35\,mas, corresponding to a
shell uniform disk diameter of $\sim$56\,mas, will provide a strong
constraint on the radii of small dust grains. However, model outputs 
are not currently available for that wavelength range.

%

\begin{figure}
\includegraphics[scale=0.7]{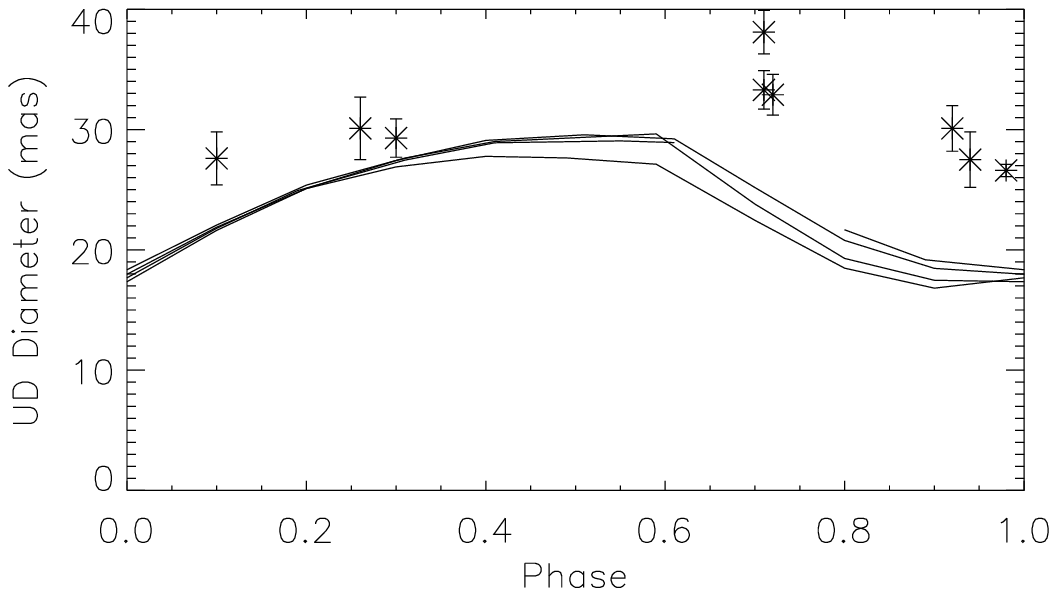} 
\caption{The same as Figure~\ref{figJDiams}, except for the H-band.}
\end{figure}

\begin{figure}
\includegraphics[scale=0.7]{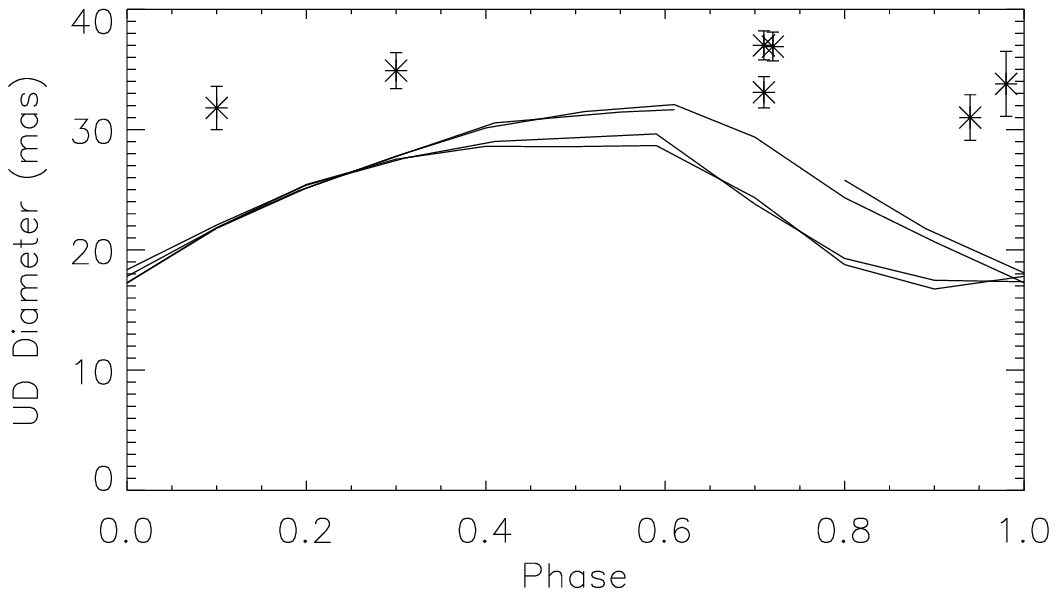} 
\caption{The same as Figure~\ref{figJDiams}, except for the K-band.}
\label{figKDiams}
\end{figure}

\begin{table*}
\caption{Summary of key observational data on $o$~Cet. Units:
  Vega magnitudes for light-curves, mJy for radio photometry,
  milli-arcsec for diameters.}
\begin{tabular}{lllllll}
\hline
Wavelength & Data Type & Observation  & Model range & Phase     & References \\
($\mu$m)   &           & Range        &             & Coverage? &  \\
\hline
0.55 &  Light Curve & 2.5--9.5 & 2.0--11& Y & AAVSO,AFOEV \\
1.2  &  Light Curve &  (-1.7)--(-0.3)& (-1.8)--(-0.6)& Y & \citet{Whitelock00} \\
1.65 &  Light Curve & (-2.6)--(-1.3)& (-2.6)--(-1.5)& Y & \citet{Whitelock00} \\
2.2  &  Light Curve & (-3.0)--(-1.8)&  (-3.0)--(-2.0)$^c$& Y & \citet{Whitelock00} \\
0.4--0.6 & Spectral Type & M5e-M9e & None$^b$ & Y & \citet{Skiff09,Samus04}\\
0.307   & Photometry  & 14.85 & None$^e$ & N & \citet{Karovska97} \\
0.346   & Diameter    & 35 (FWHM) & None$^b$ & N & \citet{Karovska97} \\
0.45-1.03 & Diameter & 31--103 & 25--60$^d$& N & \citet{Labeyrie77}\\
0.68-0.92 & Diameter & 20--60 FWHM & 16--38 & N & \citet{Ireland04a} \\
1.24    & Diameter & 22--30 & 17--34 & Y &\citet{Woodruff08}\\
2.26    & Diameter & 31--37 & 17--29 & Y &\citet{Woodruff08}\\
1.24    & Diameter & 22--30 & 17--32 & Y &\citet{Woodruff08}\\
1.1-3.8 & Diameter & 25--68 & 17--58 & N &\citet{Woodruff09}\\
11.15   &  Diameter   & 46--55 & 21--72 & Y &  \citet{Weiner03} \\
3.6\,cm$^a$ & Photometry & 0.20-0.37\,mJy  & 0.09-0.32\,mJy & N & \citet{Reid97,Matthews06} \\
7\,mm   &  Diameter & 52 & 29-55 & N & \citet{Reid07} \\
\hline
\end{tabular}\\
$^a$Shorter-wavelength radio observations are consistent with a
$\nu^2$ power law within errors. A 0.12\,mJy contribution from Mira~B
has been subtracted.\\
$^b$No model spectral type calculations are possible. See text.\\
$^c$Distance to model fixed to 107\,pc so that K-band maximum agrees.\\
$^d$When the fluorescence scattering approximation is used (as in Paper~I), 
the range becomes 26--80\,mas, with the upper diameter limited by the model 5\,$R_p$ surface.\\
$^e$Although the ultraviolet is important for deep-atmosphere temperature profiles, no ultraviolet fluxes
are output because the near-surface opaciites are likely unreliable.\\
\label{tbloCetObs}
\end{table*}

%

\section{Predicting Fundamental Parameters with Models and Observations}
\label{sectParams}

For each Mira variable modeled in this paper, there are currently 3
physical ($M$, $L$ and $Z$) parameters and 2 model parameters
($\alpha_m$ and $\alpha_\nu$). In the parameter neighborhood of the {\tt
  o54} series, we find that the radius of the ``parent'' star approximated by
linear pulsation is given by:

\begin{equation}
 \frac{R_*}{216 R_\odot} \approx (\frac{L_*}{5400 L_\odot})^{0.8} (\frac{M_*}{1.1 M_\odot})^{-0.4} (\frac{\alpha_m}{3.5})^{-0.7} (\frac{Z}{0.02})^{0.2},
\label{eqnRMixLength}
\end{equation}

or in angular units:

\begin{equation}
 \theta_* \propto d^{0.6} F_*^{0.8} M_*^{-0.4} \alpha_m^{-0.7} (\frac{Z}{0.02})^{0.2},
\end{equation}

where $d$ is the distance, and $F$ is the received
wavelength-integrated stellar flux. In this section we will only
discuss the most direct measurements of effective temperature, derived
from near-continuum interferometry and photometry, as spectral fitting in the presence of
non-LTE effects \citep{Ireland08}, metallicity and abundance errors
has yet to be demonstrated for extended M giants.

There is no dependence of the radius on $\alpha_\nu$, but there is a
small dependence of the period $P$ on $\alpha_v$ which we will neglect
here. The period, which typically is expressed in terms of mass and
radius (the so-called PMR relationship), we will express in terms of
our model parameters $L_*$, $M_*$ and $\alpha_m$:

\begin{equation}
 \frac{P_{\rm lin}}{261d} \approx (\frac{L_*}{5400
 L_\odot})^{1.8} (\frac{M_*}{1.1 M_\odot})^{-1.8} (\frac{\alpha_m}{3.5})^{-1.5} (\frac{Z}{0.02})^{0.4},
\label{eqnPeriodFormula}
\end{equation}

Although these relationships are only approximate and do not hold over
a wide range of parameters, they demonstrate the complex interplay
between the model input parameters. In principle, a measurement of
$Z$ from spectral synthesis, and measurements of period, amplitude, angular
diameter, luminosity and distance are enough to constrain $M$,
$\alpha_m$ and $\alpha_\nu$. However, a 10\% distance
uncertainty (the best of any nearby Mira) translates into a 20\% $L$ uncertainty, or a 20\% mass
uncertainty keeping everything else fixed at a given period. The relationships are further
complicated by the non-linear  
pulsation period differing significantly from the linear pulsation
period (e.g. Table~1), depending on amplitude. 

Consider first the problem of calibrating mass independently of 
pulsation models. Orbital periods for non-interacting Miras are far
too long for combined visual and spectroscopic combined orbits to
obtain dynamical masses due to the large radii of Miras. Clusters form
a potential hunting ground for Mira variables where the AGB can be
calibrated at a
known initial mass \citep[e.g.][]{Lebzelter07}, but the final mass of the Mira
is a function of the assumed mass loss history, clusters do not
easily provide the same age and metallicity range of Miras as in the
field, and a direct radius measurement is not yet possible. 
Mass can also be estimated from kinematics. This
is best done for Miras with kinematics inconsistent with the thick
disk or halo. The best example of this for nearby Miras is $o$~Cet.

$o$~Cet has a ($U$,$V$,$W$) space velocity of (-26, -62, -89)
km\,s$^{-1}$ when using the revised HIPPARCOS distance from
\citet{vanLeeuwen07}. This space velocity is unusually large for a
Mira, which is a major reason why the interaction between $o$~Cet
and the interstellar medium produces such an impressive tail
\citep{Martin07}. Although kinematics is often inconclusive when
applied to individual stars, this space velocity falls within the
97\% probability contour for thick disk membership according to
\citet{Reddy06}. In the detailed analysis of \citet{Robin03}, the
$W$ velocity of $o$~Cet is inconsistent at 5$\sigma$ with even the
old (5-10\,Gyr) thin disk, and is most consistent with being a member
of the thick disk, modeled as a
single stellar population of 11\,Gyr age. Importantly, $o$~Cet can
not be a runaway star \citep[e.g.][]{Hoogerwerf00} where its space velocity is due to a
single strong gravitational interaction in its past, because it has
retained its wide companion Mira~B, and the orbit can not be highly
eccentric as the periastron must be outside the atmosphere of $o$~Cet. 
Therefore the progenitor mass of $o$~Cet is almost certainly less than 1.1\,$M_\odot$ (e.g. the
\citet{Girardi00} evolutionary tracks places  a 1.06\,M$_\odot$ star
on the TP-AGB after 10\,Gyr, or less for sub-solar metallicity). This
places the current mass of $o$~Cet at 1\,$M_\odot$ or less. This is
close to our model value of 1.1\,$M_\odot$. However, if $L$ were to be
kept constant 
and $M$ decreased, $\alpha_m$ would have to be increased further from its
already large value in order to maintain the period. We will discuss
this further below after first discussing $\alpha_m$.

%
The comparisons of models to observations in this paper has already
provided significant evidence that model temperatures are too high,
implying that $\alpha_m=3.5$ is too high a value
for Miras with parameter ranges applicable to $o$~Cet and R~Cas. There
are 3 key pieces of evidence: the near-maximum $V$-$K$ colours, the
infrared spectral fitting for R~Cha, and the measured angular
diameters.

\begin{figure*}
\begin{minipage}[b]{0.45\linewidth}
\centering
\includegraphics[width=\linewidth]{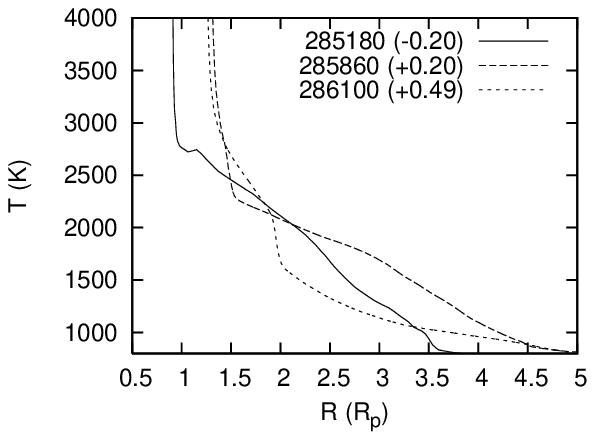}
\includegraphics[width=\linewidth]{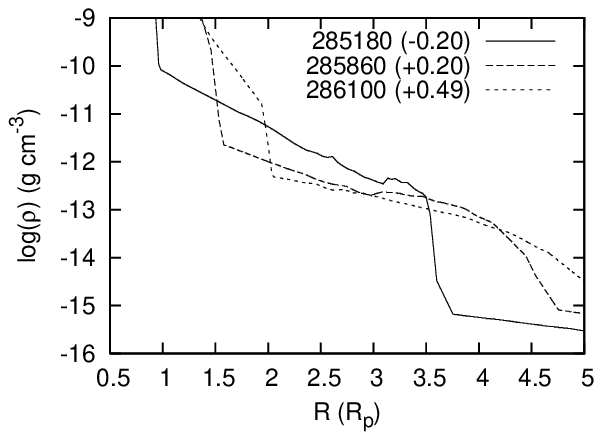}
\end{minipage}
\begin{minipage}[b]{0.45\linewidth}
\centering
\includegraphics[width=\linewidth]{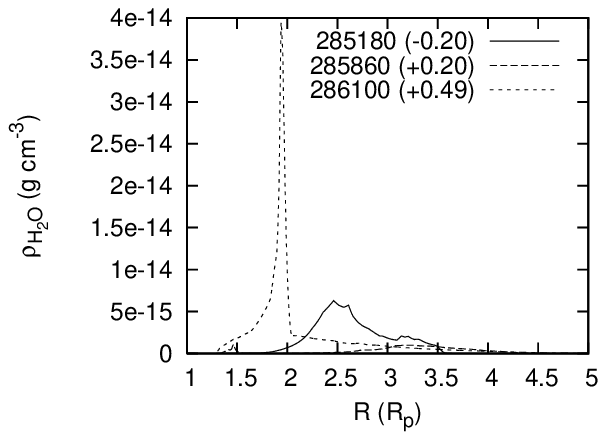}
\includegraphics[width=\linewidth]{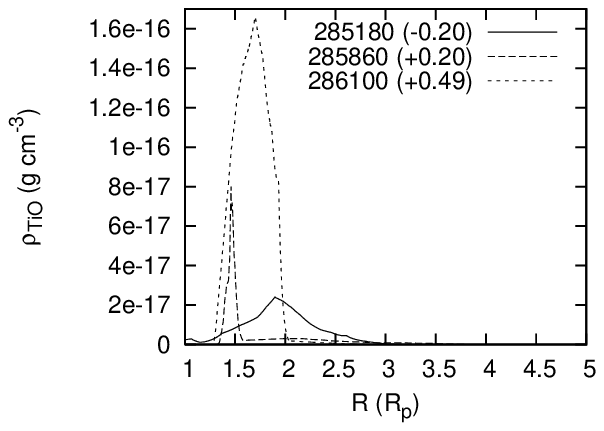}
\end{minipage}
\caption{Temperature (K ; upper left), density (g cm$^{-3}$ ; lower left panel) 
  and the moecular density $\rho_{\rm mol}$ for molecules 
  H$_{2}$O (upper right) and TiO (lower right) as a function of r/R$_p$ for 
  3 phases of a cycle of the {\tt o54} model series: 285180 (phase -0.20, 
  solid line), 285860 (+0.20, dashed), 286100 (+0.49, short-dashed).}
\label{figShell1}
\end{figure*}

\begin{figure*}
\begin{minipage}[b]{0.45\linewidth}
\centering
\includegraphics[width=\linewidth]{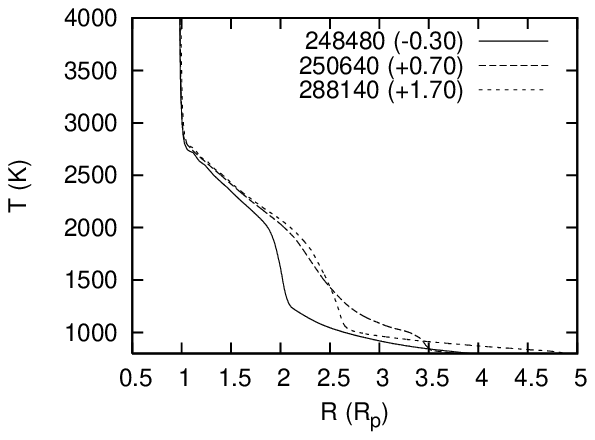}\\
\includegraphics[width=\linewidth]{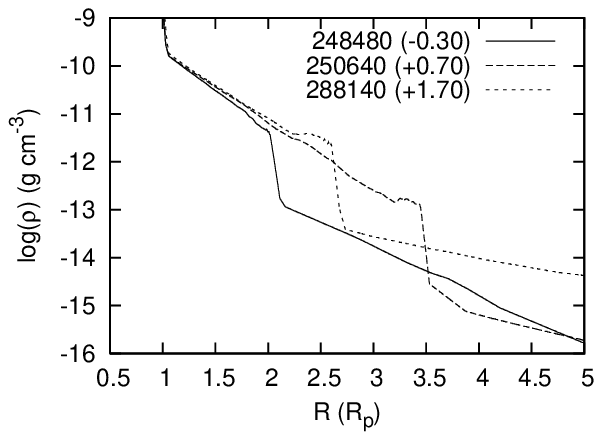}
\end{minipage}
\begin{minipage}[b]{0.45\linewidth}
\centering
\includegraphics[width=\linewidth]{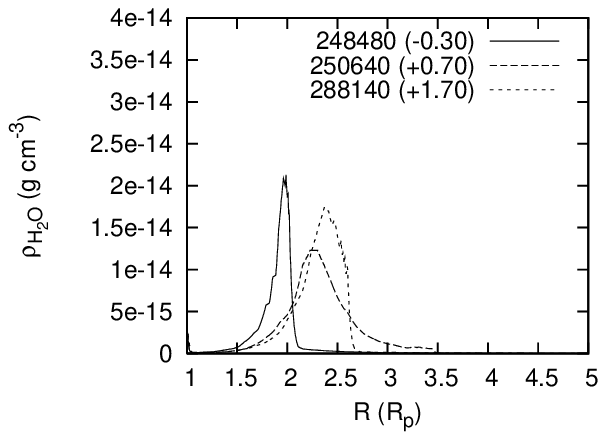}\\
\includegraphics[width=\linewidth]{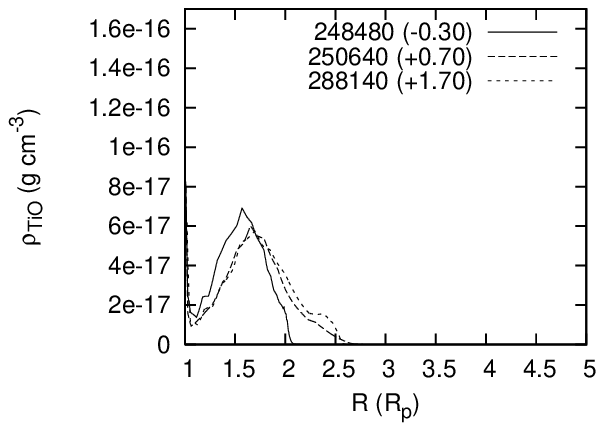}
\end{minipage}

\caption{The same as Figure~\ref{figShell1}
  for the same phase 0.7 of 3 models of different cycles of the {\tt o54} 
  model series: 248480 ("extended" series -0.30, solid line), 250640
  ("extended" series +0.70, dashed), 288140 (+1.70, short-dashed).}
\label{figShell2}
\end{figure*}

\begin{figure*}
\begin{minipage}[b]{0.45\linewidth}
\centering
\includegraphics[width=\linewidth]{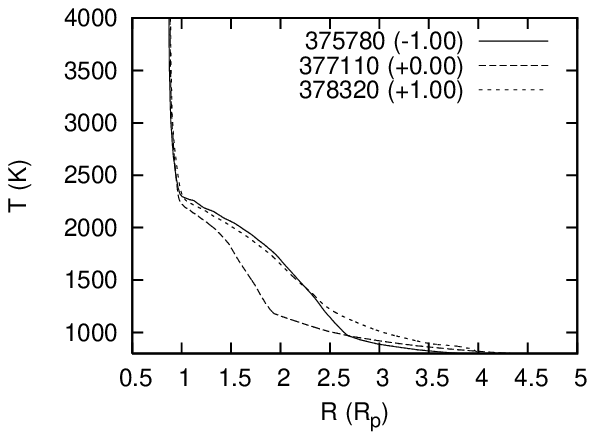}
\includegraphics[width=\linewidth]{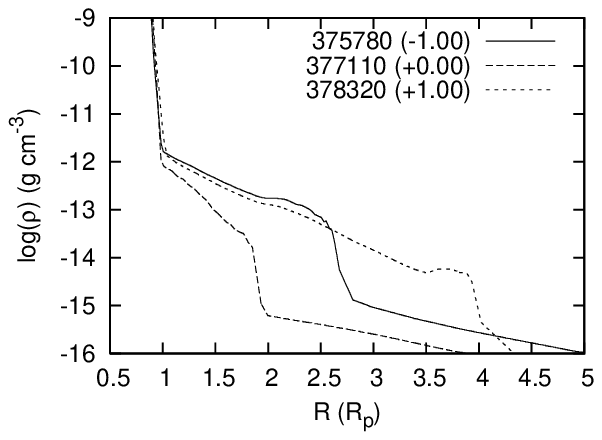}
\end{minipage}
\begin{minipage}[b]{0.45\linewidth}
\centering
\includegraphics[width=\linewidth]{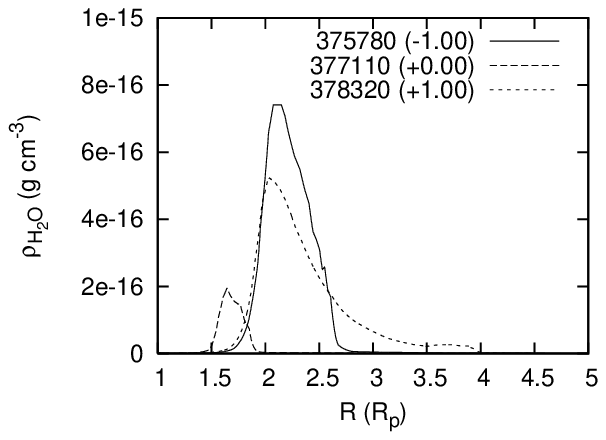}
\includegraphics[width=\linewidth]{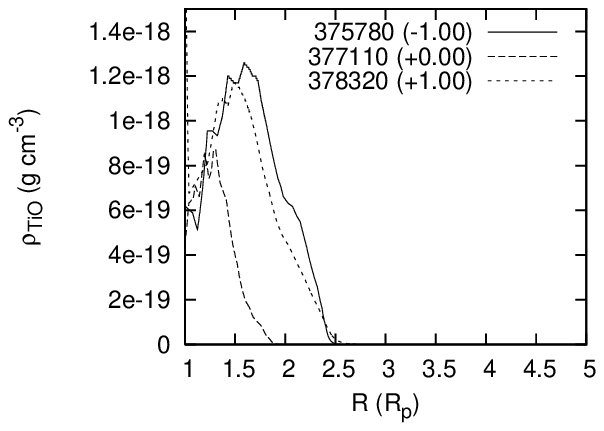}
\end{minipage}
\caption{The same as Figure~\ref{figShell1}
  for the same phase 0.0 of 3 models of successive cycles of the
  {\tt r50} model series: 375780 (-1.00, solid line), 377110 (0.00, dashed),
  3783200 (+1.00, short-dashed). }
\label{figShell3}
\end{figure*}

All series with $\alpha_m=3.5$ are too warm near
maximum, as shown by their $V$-$K$ colours of $<$5.0. The {\tt R52} and
{\tt C81} series compared to R~Leo and R~Cas respectively are the most
discrepant. Although in general this could be due to fundamental
parameters $M$ and $L$ only, in this case it is not possible. The R\,Cas {\tt
  R81} 1.35\,$M_\odot$, 8160\,$L_\odot$ model is reasonable for a
Mira: i.e. all AGB stars above a certain mass will go through an
8160\,$L_\odot$ phase, and the mass of 1.35\,$M_\odot$ at
8160\,$L_\odot$ will result from some initial mass. The systematic study
of Mira spectral type at maximum by \citet{Keenan74} found a trend
of systematically later spectral types with increasing Mira period,
with no Miras having near-maximum spectral types earlier than M5 in
the 350-500 day period range. The model spectral type of $\sim$M2 for R~Cas
is therefore much too warm. This means that the
combined choice of $\alpha_m = 3.5$ and $\alpha_\nu=0.32$ together is incorrect
for stars in the vicinity of $M=1.35$\,$M_\odot$ and
$L=8160$\,L$_\odot$. 

R~Cha is a Mira variable very much like $o$~Cet (period, amplitude,
colour), except for its more uncertain distance. In
Section~\ref{sectSpectra}, we saw that the models were $\sim$500\,K
too hot for R~Cha. However, R~Cha has the hottest 
near-maximum spectral type of all Mira variables at periods of $\sim$300--350
days in \citet{Keenan74}, so this suggests that the models are too hot
for any Mira. 
Inspection of some low-metallicity test models (to be published)
indicates that decreasing the metallicity reduces this discrepancy
but does not remove it.

Although the range of model diameters reported by \citet{Woodruff08}
were roughly consistent with the range of diameters for the Miras
studied in that paper, the minimum measured diameter for each star was
larger than the minimum diameter predicted by models. This remains
true for the current model series (see
Section~\ref{sectInterferometry}), especially for $o$~Cet. A
discrepancy in near-continuum diameters at phases $\sim$0.9-0.2, where
the photosphere is relatively compact can only be rectified with an
increase in model radius by 10--20\%. 

Let us consider how models of $o$~Cet could be modified in order to
achieve a larger apparent radius. The simplest method is to place the
star at a closer distance with a lower luminosity and the same linear
radius and mass (preserving the period).
We note that the most recent HIPPARCOS 
analysis of \citet{vanLeeuwen07} places $o$~Cet as close as 75\,pc
within 2$\sigma$, which would mean luminosities as low as
2650\,$L_\odot$ are consistent with direct observations. Such a low
luminosity would, however, be 0.8 magnitudes below the LMC P-L
relationship \citep{Whitelock08}. This relationship provides a
stronger constraint on the absolute K-magnitude of $o$~Cet, as the
dispersion in the relationship 
is only 0.13\,magnitudes \citep{Feast89}. Adopting this relationship places
a 2$\sigma$ lower limit on the $o$~Cet luminosity of
4400\,$L_\odot$. By solving the approximate equations
\ref{eqnRMixLength} through \ref{eqnPeriodFormula} after decreasing the
$M$ to 1\,M$_\odot$ and increasing angular diameter by 10\%, the model
luminosity becomes 4200\,$L_\odot$ and the mixing length parameter
$\alpha_m=2.9$ (effects of $\alpha_\nu$ and $Z$ are neglected here).  
It is therefore likely not possible to fit all
observations of $o$~Cet unless the model mass is greater than
1\,$M_\odot$ or the luminosity more than 2$\sigma$ below the LMC P-L
relationship -- each of which would be a controversial claim. This discussion
shows just how difficult it is to provide precisely calibrated models
of Mira variables without clear observational reference points.

The key problem here of finding the best value for $\alpha_m$ as a
function of $M$ and $L$ can be expressed as a problem of
finding the radii of real stars at given $M$ and $L$. Mixing length
only provides a way to calculate $R$ for given $M$ and $L$ once free 
parameters (especially the mixing length) are fixed based on known stars. 
Models of main-sequence stars and even
K-giants can be calibrated very well from the sun and other stars of
accurately known parallax. However, extrapolating to M giants, where
pressure scale-heights can be a significant fraction of a radius, is
not expected to be reliable. 

What is therefore needed is a sample of M giants of well-known $M$,
$L$ and $R$, from which to calibrate the mixing length. Although we
can not measure $M$, period can serve as a proxy for $M$ given $L$ and
$R$. $L$ and $R$ can be measured accurately for a sample of stars with
relatively compact atmospheres (i.e. with a well-defined $R$) that have a well known
parallax, photometry and effective temperature. At this 
point, spectral synthesis is not reliable enough or calibrated well
enough for M giants to produce accurate effective temperatures and
compositions. Therefore, this sample should include semi-regular
pulsators with accurate photometry, angular diameters and
periods. Either accurate periods or angular diameters are currently
missing for many of the closest M giants, so we suggest that measuring
and collating such information should be an active area of research.

Assuming that a reasonable value for $\alpha_m$ can be prescribed for
Mira models, the main free parameters for any individual Mira are
composition and $\alpha_\nu$. Composition (primordial metallicity and
C/O ratio) must come from spectral synthesis.  Given a set of Miras
with well-measured distances, $L$ is determined from observations and
$M$ from the period of the Mira - only $\alpha_\nu$ can be used to tune
the model amplitude and should be relatively easy to calibrate.



\section{The Molecular-Shell Scenario}
\label{sectShell}

Interferometric measurements of Mira variables have been shown to be
internally consistent only if there is a layer of molecular water far
above the continuum-forming photosphere \citep[e.g.][]{Weiner04}. 
Observational comparison with models that
include this kind of water and/or dust layer have so-far been
dominated by non-physical models, i.e. those that neither provide a
mechanism for elevating the emitting material nor calculations of the
chemistry that determines which components dominate the radiative
transfer at which radii.

Nevertheless, these ad-hoc models have provided a relatively simple
picture for the regions around Mira variables and have impressively
fitted a limited selection of observable properties. In this section
we will examine the physical and observable properties of the
molecular shells in our model series.

Figs. 1 to 4 show that cycle-to-cycle variations in the pulsation models are
generally quite modest in terms of the luminosity, of the temperature at and
the position of the $\tau_g$=2/3 layer and of the position of deep-layer mass 
zones below the $\tau_g$=2/3 layer. The o54 and C81 model series show 
somewhat more pronounced cycle-to-cycle effects than the r52 and C50 models. 
Inspection of specific numerical values of the radius $R$ of the 
$\tau_{\rm Ross}$=1 layer and the therefrom derived effective temperature 
$T_{\rm eff} \propto (L/R^{2})^{1/4}$ given in Tabs. 2 to 8 for the non-grey 
atmospheric stratifications confirm this cycle stability.

In contrast, we notice substantial differences between 
different cycles, and often between successive
cycles, in terms of the positions of high-layer mass zones (Figs. 1 to 4). Ê
These differences are closely related to substantial differences of the 
strengths and positions of outward traveling shock fronts. Inspection of 
shock-front positions (Tabs. 2 to 8) in the selected cycles for which 
detailed atmospheric models were computed show a shock front typically 
emerging at pre-maximum phase around -0.3 to -0.1, then traveling outward 
during about 1 to 1 1/2 cycles while it becomes weaker and slower before 
the subsequent front catches up and both fronts merge. Typically, the outer 
front starts retreating before merger, but occasional shock fronts 
traveling towards circumstellar space (where they eventually fade away) 
are also seen in the Tables. 

The positions and heights of shock fronts at different phases and in 
different cycles determine the upper atmospheric density stratification and, 
therefrom, the details of the temperature stratification and of the 
partial-pressure stratification of molecular absorbers. The
assumptions of local thermodynamic equilibrium and spherical symmetry
are sufficient to derive 
this stratiÞcation (as may not be the case
for the so-called MOLsphere \citep{Tsuji00} in supergiants, due to the
co-existence of the chromosphere in those stars).
The study of \citet{Tej03b}, based on models of \citet{Hofmann98}, 
shows that the details of shock-front propagation may lead to strong 
cycle-to-cycle differences of the stratification of the outer atmosphere
resulting in strong differences of the density and geometric characteristics
of water ``shells'', i.e. of layers whose absorption is dominated by water
molecules. 

Figures~\ref{figShell1} to \ref{figShell3} demonstrate, for the
{\tt o54}
and the {\tt r50} series of the here
presented {\small CODEX} model sets, the drastic phase and cycle effects of
shock-front propagation on the temperature-density stratification and
on the appearance of H$_{2}$O and TiO ``shells''. Density decreases
monotonically with radius,
while the sharp decrease at a  shock front provides
the {\em outer} edge of any ``shell''. The decrease
of temperature with radius provides a relatively sharp edge to the
region where water can exist in chemical equilibrium, and this
provides the {\em inner} edge of any ``shell''. At most infrared wavelengths,
gas is reasonably transparent between temperatures where H- opacity is
dominant ($\ga 3000$\,K) and where water is dominant ($\la 1800$\,K).
A similar pattern is found for the TiO molecule which, however, is formed
in somewhat deeper layers than water and, therefore, does not depend so
strongly on upper-atmosphere shock-fronts and shows smaller, though
by no means negligible, cycle-to-cycle effects.

\begin{figure}
\hspace{-0.5cm}\includegraphics[width=1.08\linewidth]{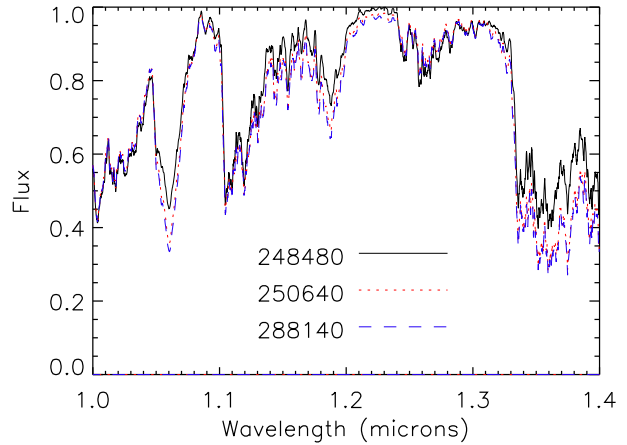}
\caption{Spectra of the 3 same-phase models of Figure~\ref{figShell2}
  showing cycle-to-cycle differences of the spectrum in the J-band region of the spectrum.} 
\label{figShell4}
\end{figure}

The model-predicted effects of such molecular ``shells''
upon the absorption properties of the stellar atmosphere at observationally
important wavelengths have been discussed and compared to observations by
\citet{Ireland08}, \citet{Woodruff09} and Wittkowski et al. (in preparation).
Figure~\ref{figShell4} shows typical cycle-to-cycle differences
seen in the water-contaminated spectrum  of the 3 models presented 
in Figure~\ref{figShell2}.
Typical effects of water ``shells'' upon the shape of the centre-to-limb
variation have been discussed in the model study of \citet{Tej03b}.

Though semi-empirical ``shell'' scenarios have been used with remarkable 
success for interpreting spectroscopic and interferometric observations 
of absorption features of H$_{2}$O
\citep[e.g][]{Matsuura02,Mennesson02,Ohnaka04a,Perrin04,Weiner04} 
and of TiO \citep[e.g][]{Reid02}, models show that such scenarios can 
at best provide a very rough picture of the 
approximate instantaneous position and extent of absorbing
layers. ÊSuch semi-empirical models cannot provide any
information on changes of these layers with phase
(since local molecule abundance and resulting molecule
absorption as a function of local values of $\rho (r)$ and $T(r)$ sensitively
depends on details of shock-front progression). Note also that 
absorption by CO in low excitation lines extends from the
continuum-forming photosphere at $\sim$3000\,K right to the wind
region, so a CO ``shell'' scenario \citep[e.g.][]{Mennesson02} should be 
considered with particular caution.

Finally, we note that the existence of shell-like structures
noticeably change the computed spectra in the models,
but the effects are not so strong that the
detailed shell structure can be directly inferred from low spectral 
resolution observations. Modeling high spectral resolution
observations is beyond the scope of this paper, but such a study 
would have to take into account the velocity structure of the
atmosphere explicitly \citep[e.g][]{Nowotny10}.


\section{Radiative Acceleration and Mass Loss}
\label{sectRadAccel}

As described in \citet{Hofner08}, the conditions for radiative
acceleration to drive mass loss are that dust must be able to form,
and that the opacity exceeds the critical opacity:

\begin{equation}
\kappa > \frac{4\pi c G M_*}{L_*} = 12830{\rm \,cm^2g^{-1}} (\frac{M_*}{M_\odot})(\frac{L_*}{L_\odot})^{-1}.
\end{equation}

This opacity is attainable at solar metallicity 
with fully-condensed iron-rich dust, which
is not stable until approximately 5 continuum stellar radii
\citep{Woitke06}. This opacity is also reached by forsterite
(i.e. Fe-poor silicate) grains of 400\,nm radius due to strong
scattering \citep{Hofner08}.

The {\small CODEX} models have a chemical equilibrium model of dust
formation, so are not appropriate for modeling the slowly-growing Fe-rich
dust at $\la $5\,R$_p$, and indeed we artificially cut-off Si
condensation at a condensation fraction of 0.25 for this
reason. However, as discussed in \citet{Ireland06}, the prescription
we use for dust formation is reasonably accurate for Fe-poor silicates.

\begin{figure}
\hspace{-0.5cm}\includegraphics[scale=0.65]{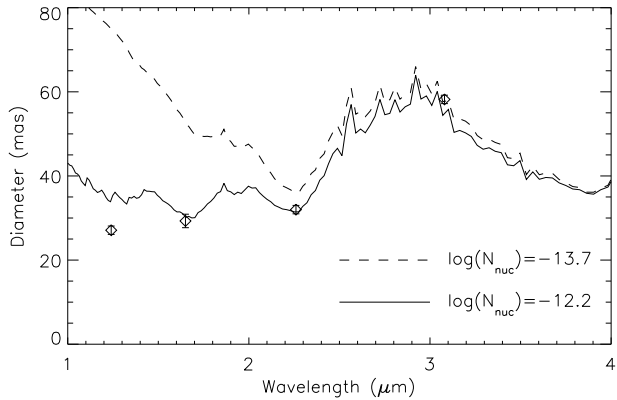}
\caption{Interferometric diameters for the {\tt o54} series model
  286060 (phase 0.41), fit to the spatial frequency where visibility
  $V=0.5$. The 
  solid line corresponds to the default number of dust nuclei, and the
solid red line corresponds to a factor of 3.1 increase
in grain radius, at log($N_{\rm nuc}$)=-13.7. Observations from 
\citet{Woodruff08} at phase 0.3 are over-plotted. Despite still not being
able to drive a wind, the 
large diameters at short wavelengths are clearly inconsistent with
observations.} 
\label{figLargeDust}
\end{figure}

The strongest observational constraints on the radii of dominant dust
species are observations that probe the opacity at short wavelengths where
dust scattering is dominant, and wavelengths where water absorption is
dominant. As optically thin scattering does not affect the spectrum,
the best observations to probe this difference are resolved
observations of Miras as a function of
wavelength. Figure~\ref{figLargeDust} shows the interferometric diameters
of $o$~Cet as a function of wavelength as measured by
\citet{Woodruff08} along with the diameters predicted by the 286060 model
and the diameters predicted by the same model with the base-10
logarithm of the number of dust nucleii per H atom $N_{\rm nuc}$ decreased from
-12.2 to -13.7, and the corresponding maximum grain radius increased
from 63 to 194\,mn. These large grains still have insufficient opacity
to overcome gravity in this model. We could not increase the grain radii further
without the optical depths becoming too large at our chosen 5\,$R_p$
surface, and the preferred value from \citet{Hofner08} of log($N_{\rm
  nuc}$=-15) would produce diameters that are far too large. 
It is clear that these large grain radii are not
consistent with the relatively large H$_2$O column densities at
several continuum radii inferred from infrared interferometric
observations. Either radiation pressure on small Fe-rich grains or
large Fe-poor grains could drive winds from M-type Mira variables, but
the base of the wind and the grains that drive it must originate from
layers higher than where H$_2$O is seen in the near-infrared,
i.e. higher than about 3--4 continuum-forming radii.

\section{Conclusions and Future Work}
\label{sectConclusions}

The atmospheric models presented here, based on self-excited pulsation
models and opacity-sampling treatment of radiation transport, provide a
fairly realistic approximation of the atmospheric density-temperature
stratification. Many spectral features are predicted with satisfactory
accuracy but some, like TiO bands, require further improvement of the
models e.g. the non-LTE treatment of Paper~I. The present sequence of models 
also comprises only 4 combinations of 
basic stellar parameters at only a single composition and, therefore, can only describe a relatively
small subset of Mira variables. Predictions of the 
4 model series presented here (Table 1) are available on-line (Section~\ref{sectUsingModels}).

There are several potential causes of substantial deviation from spherical asymmetry in 
Mira variables -
including convective cells, weak chaos and Rayleigh-Taylor instabilities \citet{Woitke06}. 
Asymmetries are clearly not considered in our model series as they are spherically symmetric, 
but asymmetries are relatively common in Mira variables when observed at sufficient angular 
resolution (e.g. \citet{Ragland06}). The one prediction that can be made from the models is that 
the wavelengths most susceptible to cycle-to-cycle variations (e.g. L-band where the water shells 
are optically-thick) should also show asymmetries, as high layers on opposite sides of the star 
should not be strongly causally connected and show weak chaos. We finally suggest that an 
observation so far missing in the literature is the astrometric motions of the radio photosphere over 
several cycles, which should be a strong indicator for the degree of high-layer chaos in 
Mira atmospheres.

Determination of the internal fundamental model parameters, i.e. mixing-length
$\alpha_m$ and turbulent viscosity $\alpha_\nu$ (Table 1), would require
observation of a set of stars with different mass, luminosity and pulsation
period.
For a given pulsation period, a higher-mass star must have a 
higher luminosity (radius),
but a similar effective temperature. We suggest that further studies
of low-amplitude pulsators with Mira-like periods such as R~Dor and
W~Hya may provide the key to tuning the mixing length parameter
$\alpha_m$ of Mira model  
series. These are likely higher-mass stars, but with a similar effective
temperatures to $o$~Cet. Tuning the turbulent viscosity parameter will
best be done by fitting to amplitudes of models with the best-known masses
such as, e.g., those kinematically associated with the thick disk.

\section*{Acknowledgments}
M.I. would like to acknowledge support from the Australian Research
Council through an Australian Postdoctoral Fellowship. PRW was partially 
supported by Australia Research Council
grant DP1095368. We acknowledge
with thanks the variable star observations from the AAVSO
International Database contributed by observers worldwide and used in
this research. This research has made use of the AFOEV database,
operated at CDS, France.


\label{lastpage}
\end{document}